\documentclass[bibyear]{aa}
\usepackage{graphicx,url,cancel}
\usepackage{mathrsfs,amssymb,amsmath}
\usepackage{subfigure}

\usepackage[varg]{txfonts}

\newcommand{\ud}{\mathrm{d}}
\newcommand{\be}{\begin{equation}}
\newcommand{\ee}{\end{equation}}

\newcommand{\h}{{\rm H}}
\newcommand{\hh}{{\rm H_{2}}}
\newcommand{\he}{{\rm He}}

\begin{document}

\title{Cosmic-ray ionisation in circumstellar discs}

\author{Marco~Padovani\inst{1}, Alexei~V.~Ivlev\inst{2}, Daniele~Galli\inst{1} and Paola~Caselli\inst{2}}

\authorrunning{M. Padovani et al.}

\institute{
INAF--Osservatorio Astrofisico di Arcetri, Largo E. Fermi 5, 50125
Firenze, Italy\\
\email{[padovani,galli]@arcetri.astro.it}
\and
Max-Planck-Institut
f\"ur extraterrestrische Physik, Giessenbachstr. 1, 85748 Garching, Germany\\
\email{[ivlev,caselli]@mpe.mpg.de}
}


\abstract
{Galactic cosmic rays (CRs) are a ubiquitous source of ionisation of the interstellar gas, competing with UV and X-ray
photons as well as natural radioactivity in determining the fractional abundance of electrons, ions, and charged dust grains
in molecular clouds and circumstellar discs.}
{We model the propagation of various components of Galactic CRs versus the 
column density of the gas. Our study is focussed on the propagation at high densities, above a few g~cm$^{-2}$, especially
relevant for the inner regions of collapsing clouds and circumstellar discs.}
{The propagation of primary and secondary CR particles (protons and heavier nuclei, electrons, positrons, and photons)
is computed in the continuous slowing down approximation, diffusion approximation, or catastrophic approximation by
adopting a matching procedure for the various transport regimes. A choice of the proper regime depends on the nature
of the dominant loss process modelled as continuous or catastrophic.}
{The CR ionisation rate is determined by CR protons and their secondary electrons below $\approx 130$~g~cm$^{-2}$ and by electron-positron pairs created by photon decay above
$\approx600$~g~cm$^{-2}$. We show that a proper description of the
particle transport is essential to compute the ionisation rate in the latter case, since the 
electron and positron differential
fluxes depend sensitively on the fluxes of both protons and photons.}
{Our results show that the CR ionisation rate in high-density environments, such as the inner parts of collapsing
molecular clouds or the mid-plane of circumstellar discs, is higher than previously assumed. It does not decline
exponentially with increasing column density, but follows a more complex behaviour because of the interplay of
the different processes governing the generation and propagation of secondary particles.}


\maketitle

\section{Introduction}

Ionisation plays a crucial role in cold, dense molecular cloud cores and in circumstellar discs, where it controls the
chemical properties of the gas, the coupling to the magnetic field, and (in the latter case) the generation of turbulence
via the magnetorotational instability. The primary agents of ionisation in dense gas at column densities much larger than the
values at which interstellar (IS) UV photons are absorbed (i.e. at visual extinctions $A_{\rm V}\gtrsim 1$~mag or column
densities $N\gtrsim 10^{21}$~cm$^{-2}$) are X-rays, cosmic rays (CRs), and decaying of radionuclides. Deep in the densest
parts of molecular clouds, characterised by densities $n\gtrsim 10^5$--$10^6$~cm$^{-3}$ and
$A_{\rm
V}\gtrsim 50-100$~mag, CR ionisation is dominant, leading to an ionisation fraction decreasing with density
(McKee~\cite{m89}; Caselli et al.~\cite{c02}; Walmsley et al.~\cite{w04}; Maret et al.~\cite{m06}). In discs around young,
active stars, or in molecular clouds close to supernova remnants or massive star-forming regions, the situation is
complicated by the effects of stellar X-rays (Glassgold et al.~\cite{gn97}), the possible exclusion of low-energy CRs by
stellar winds (Cleeves et al.~\cite{ca13a}), and the presence of local sources of accelerated particles (Lee et
al.~\cite{ls98}; Padovani et al.~\cite{phm15,pmh16}). The efficiency of stellar X-rays to ionise the circumstellar gas
depends on total fluxes and hardness of the spectra. For example, hard X-rays of energies 1, 6 and 20~keV are absorbed by
column densities of $2.5\times 10^{22}$, $4.5\times 10^{24}$ and $1.3\times 10^{25}$~cm$^{-2}$, respectively (Igea \&
Glassgold~\cite{ig99}).

In the shielded regions near to the disc mid-plane, where X-rays and CRs are strongly attenuated, radioactive elements may
substantially contribute to the electron fraction. In this case the ionisation rate is proportional to the abundance of the
radioactive element and its decay rate. Thus, short-lived radionuclides (SLRs; mostly $^{26}$Al with half-life $7.4\times
10^5$~yr) contribute comparatively more than long-lived radionuclides (LLRs; mostly $^{40}$K with half-life $1.3\times
10^9$~yr), but decay faster. Assuming the $^{26}$Al abundance inferred for the early solar system from the analysis of
meteorites (McPherson et al.~\cite{md95}; Umebayashi \& Nakano~\cite{un09}), SLRs can maintain an ionisation rate of
$\approx 10^{-19}$~s$^{-1}$ in the mid-plane of a disc for $\approx 1$~Myr, while LLRs lead to ionisation rates of only
$\approx 10^{-22}$~s$^{-1}$, but on timescales much longer than the disc lifetimes (Umebayashi \& Nakano~\cite{un09};
Cleeves et al.~\cite{ca13a}). However, since the average Galactic abundance of $^{26}$Al inferred from its $\gamma$-ray
emission is about one order of magnitude lower than the meteoritic solar system value (Diehl et al.~\cite{dh06}), the
probability for a star of being born in a $^{26}$Al-rich environment, as in the case of the Sun, is relatively low
(Gounelle~\cite{g15}).

The propagation of low-energy CRs at low column densities, which is characteristic of the diffuse envelopes of molecular
clouds, is mostly determined by resonant scattering on self-generated 
magnetohydrodynamic (MHD) waves (occurring on the scale of the particle
gyroradius; see e.g. Cesarsky \& V\"olk~\cite{cw78}; Everett \& Zweibel~\cite{ez11}; Morlino \& Gabici~\cite{mg15}; Ivlev
et al.~\cite{ivl18}). However, at gas densities higher than $\approx 10^3-10^4$~cm$^{-3}$ such waves cannot affect the CR
transport because they are completely damped from efficient ion-neutral collisions (Ivlev et al.~\cite{ivl18}).

At any given depth in a cloud or circumstellar disc, CRs are attenuated in a way that generally depends on
characteristics of the ambient medium. In a cloud threaded by a large-scale magnetic field, CRs perform helical
trajectories along the local field lines, i.e. CRs gyrate many times before they collide with a particle of the
medium. Therefore, if the field lines are strongly twisted, the effective column density seen by a CR particle at
a given point can be much larger than the line-of-sight column density at that point (Padovani \& Galli~\cite{pg11,pg13};
Padovani et al.~\cite{phg13}). The decrease of the CR ionisation rate follows a power-law behaviour as function of the
effective column density, $N$, in the range $10^{20}-10^{24}~{\rm cm^{-2}}$, corresponding to effective surface densities,
$\Sigma$, below a few ${\rm g~cm^{-2}}$ (Padovani et al.~\cite{pgg09}, hereafter PGG09). At higher densities, the
attenuation is generally assumed to be exponential with a characteristic scale of $\approx 96$~g~cm$^{-2}$ (Umebayashi \&
Nakano~\cite{un81}).

The main goal of this paper is to determine the attenuation of CRs at moderate-to-high surface densities (larger than a few
g~cm$^{-2}$) by including both the energy loss and the particle production mechanisms relevant for the inner regions of
circumstellar discs or collapsing clouds and adopting appropriate models for the transport of primary and secondary CR
particles. We show that above $\approx130$~g~cm$^{-2}$ the CR ionisation rate 
rapidly becomes dominated by electron-positron
pairs that are locally produced by secondary photons. As the latter are insensitive to the magnetic field and propagate
isotropically, above this threshold the ionisation is controlled by the line-of-sight (rather than the effective)
column density.

The paper is organised as follows: In Section~\ref{spectra} we summarise two reference models for the IS
spectra of CR protons and electrons; in Section~\ref{energylosses} we examine the dominant energy loss
mechanisms for primary and secondary CR particles operating at different energies; in Section~\ref{propagation} we describe
our modelling of the propagation and attenuation of primary CRs; the generation and propagation of secondary particles is
described in Section~\ref{generation}; in Section~\ref{ionisation} we compute the total CR ionisation rate, focussing on high
column densities; in Section~\ref{discussion} we discuss implications of our results for various astrophysical
environments; and in Section~\ref{conclusions} we summarise our most important findings.

\section{Interstellar CR spectra}
\label{spectra}

The IS differential fluxes of CR nuclei (hereafter, IS CR spectra)
at high energies, $E\gtrsim 1$~GeV/nuc, 
are well
constrained by various sets of observations from ground-based to balloon and satellite detectors (e.g. Aguilar et
al.~\cite{aa14,aa15}). On the other hand, low-energy IS 
spectra, being strongly affected by solar modulation effects (e.g.
Putze et al.~\cite{pm11}), cannot be reliably determined by the same means. The best available estimate of the spectra
(both nuclei and electrons) at energies $E\lesssim 500$~MeV/nuc is provided by the most recent Voyager~1 observations
(Cummings et al.~\cite{cs16}), down to $\approx 1$~MeV/nuc and $\approx 10$~MeV for CR nuclei and electrons, respectively.

In this paper we adopt the analytical expression for the IS spectra of
CR electrons and protons\footnote{IS spectra of heavier nuclei (of given abundances)
are also described by Eq.~(\ref{jis}).}, as described in
Ivlev et al.~(\cite{ip15}),
\be\label{jis}
j_{k}^{\rm IS}(E)=C\frac{E^{a}}{(E+E_{0})^{b}}~\mathrm{eV^{-1}~s^{-1}~cm^{-2}~sr^{-1}}\,,
\ee
where we slightly modify the values of the parameters $E_{0}$ and $a$, to better 
reproduce the most recent Voyager~1 data release (see Table~\ref{tab:jis}).
A simple extrapolation of the Voyager~1 data at lower energies fails to reproduce the CR ionisation rate measured in diffuse
clouds from H$_3^+$ emission (Indriolo \& McCall~\cite{im12}). For this reason, 
we consider two different models for the CR proton spectrum: 
a ``low'' spectrum, ${\mathscr L}$,
obtained by extrapolating the Voyager~1 data ($a=0.1$), and a ``high'' spectrum, ${\mathscr H}$ 
($a=-0.8$). The resulting ionisation rates and their comparison with observational data are discussed in Ivlev et
al.~(\cite{ip15}). The ${\mathscr L}$ and ${\mathscr H}$ proton spectra must be considered as lower and upper bounds,
respectively, to the actual average Galactic CR spectrum. Although it is generally assumed that the CR intensity
measured by Voyager~1 spacecraft is not affected by the solar wind modulation, one should not forget that the observed
magnetic field has not changed yet to the direction expected in the IS medium (Burlaga et al.~\cite{bn13}),
suggesting the possibility that the spacecraft may reside in a region of compressed solar wind (Fisk \&
Gloeckler~\cite{fg14}; Gloeckler \& Fisk~\cite{gf15}). Thus, care should be taken in interpreting the Voyager~1 measurements
as representative of the Galactic spectrum.

\begin{table}[!h]
\caption{Parameters of IS CR spectra, Eq.~(\ref{jis}).}
\begin{center}
\begin{tabular}{lcccc}
\hline\hline
Species ($k$) & $C$ & $E_{0}/\rm{MeV}$ & $a$ & $b-a$\\
\hline
$e$ & $2.1\times10^{18}$ & 710 & $-1.3$ & 3.2\\
$p$ (model $\mathscr{L}$) & $2.4\times10^{15}$ & 650 & $\phantom{-}$0.1 & 2.7\\
$p$ (model $\mathscr{H}$) & $2.4\times10^{15}$ & 650 & $-0.8$ & 2.7\\
\hline
\end{tabular}
\end{center}
\label{tab:jis}
\end{table}%

In Sect.~\ref{propagation} (see Fig.~\ref{pspectra}), we show that at high column densities,
typical of circumstellar discs, the propagated CR proton spectrum 
becomes independent of the assumptions on the slope at low energy.

\section{Energy losses and attenuation of CRs}\label{energylosses}

In order to calculate the ionisation induced by CRs in molecular clouds or circumstellar discs, we need to study the
propagation and attenuation of CR species $k$ (including secondaries), and derive their spectra $j_{k}(E,N)$ as function of
the column density, $N$, along the direction of propagation, i.e. along the local magnetic field. We consider the
propagation of CRs in a semi-infinite medium and, hence, assume that half of IS CRs (with the
energy spectra described in Sect.~\ref{spectra} and isotropic pitch-angle distribution) are incident on the surface. The inclination of the magnetic field with
respect to the surface can be arbitrary.
To simplify the
presentation, we first obtain results for zero pitch angle and compute the effect of the angle averaging on the ionisation
in Sect.~\ref{ionisation}. For applications to a circumstellar disc, in Sect.~\ref{discussion} we consider CRs coming from
both sides of the disc. 

We assume that hydrogen is only in molecular form and use the IS medium composition by
Wilms et al.~(\cite{wa00}), summarised in Table~\ref{tabwilms}. The mean molecular weight of the medium, $\bar A$, is
\be \bar A=\sum_{Z}A_{Z}f_{Z}=2.35\,, \ee
where $A_{Z}$ and $f_{Z}$ are the mass number and abundance with respect to the total number of particles,
respectively.\footnote{For a solar composition (e.g. Anders \& Grevesse~\cite{ag89}), the mean molecular weight and
resulting total ionisation rate (see Sect.~\ref{ionisation})
vary by less than 2\%.} 
The column density is related to the surface density, $\Sigma=\bar A m_{p}N$, where $m_{p}$ is the proton mass.
Numerically, the relation between the $N$ and $\Sigma$ is given by
\be\label{NSigma}
\frac{N}{\rm cm^{-2}}=2.55\times10^{23}\frac{\Sigma}{\rm g~cm^{-2}}\,.
\ee

In order to evaluate the total CR ionisation rate, a careful treatment of showers of secondary species (photons,
electrons, and positrons) produced by primary CRs through processes such as pion decay, bremsstrahlung
(BS), and pair production
is required. In the following subsections we describe different energy loss processes for each CR species interacting with
particles of a medium of given composition.

Throughout this paper, subscripts and superscripts denote CR species and the interaction processes,
respectively. The expression for the partial energy loss function, $L_k^{l}$, for a particle of species $k$ depends on the
type of process $l$: if only a small fraction of the particle kinetic energy is lost in each collision with a particle of
the medium, the process can be considered as continuous and described by the loss function
\be\label{loss_cont} L_k^l(E) = \int_{0}^{E^{\rm max}}E'\frac{\ud\sigma_k^l(E,E^{\prime})}{\ud E^{\prime}}\ud E^{\prime}\,,
\ee
where $\ud\sigma_k^l/\ud E^{\prime}$ is the differential cross section of the process and $E^{\rm max}$ is the maximum
energy lost in a collision (see e.g. Appendix~\ref{Csigma} for Compton scattering). In the extreme case, where the entire kinetic energy is lost in
a single collision or the CR particle ceases to exist after the collision, the process is called  
catastrophic and
the loss function becomes
\be\label{loss_cat} L_k^l(E) = E\sigma_k^l(E)\,, \ee
where $\sigma_k^l$ is the cross section of the process.
Where possible, we express the total energy loss function $L_k=\sum_lL_k^l$
in terms of the loss functions for collisions with atomic hydrogen ($L_{k,\h}$) or helium ($L_{k,\he}$).

\subsection{Protons}

The proton energy loss function, $L_{p}$, is composed by two terms: ionisation losses at low energies
($L_{p}^{\rm ion}$)  
and losses due to pion production above the threshold energy $E^{\pi}$=280~MeV
($L_{p}^{\pi}$). Therefore,
\be\label{lossprotons} 
L_{p}(E)=\varepsilon^{\rm ion}L_{p,\h}^{\rm ion}(E)
+\varepsilon^{\pi}L_{p,\h}^{\pi}(E)\,, 
\ee
where $\varepsilon^{\rm ion}=2.01$
and $\varepsilon^{\pi}=2.17$ account for the presence of
heavy elements in the target medium. The two terms on the right-hand side of Eq.~(\ref{lossprotons}) are described in detail in
Appendix~\ref{Lproton}.

\subsection{Electrons and positrons}

The electron and positron energy loss function, $L_{e}$, has contributions due to ionisation losses at low
energies ($L_{e}^{\rm ion}$), BS above $\approx100$~MeV ($L_{e}^{\rm BS}$), and synchrotron above $E^{\rm
syn}\approx1$~TeV ($L_{e}^{\rm syn}$). Then, $L_{e}$ is given by
\be\label{losselectrons} L_{e}(E)=\varepsilon^{\rm ion}L_{e,\h}^{\rm ion}(E)+\varepsilon^{\rm BS}L_{e,\h}^{\rm
BS}(E)+L_{e}^{\rm syn}(E)\,, \ee
where the factors, $\varepsilon^{\rm ion}=2.01$ and $\varepsilon^{\rm BS}=2.24$, are described in Appendix~\ref{Lelectron}.
We note that the synchrotron loss term in Eq.~(\ref{losselectrons}) does not depend on the medium composition.

\subsection{Photons}

Photons are generated through BS by electrons and positrons and through decay of neutral pions (produced
by CR protons). In Sections~\ref{generation} and~\ref{ionisation} we show that the reverse process of electron-positron pair
production by photons is crucial for ionisation at high $N$, so photon propagation should be carefully treated. The
photon energy loss function, $L_{\gamma}$, is a sum of three terms: photoionisation ($L^{\rm PI}_{\gamma}$), Compton
scattering ($L_{\gamma}^{\rm C}$), and pair production losses ($L_{\gamma}^{\rm pair}$),
\be\label{lossphotons}
L_{\gamma}(E)=L^{\rm PI}_{\gamma}(E)+\varepsilon^{\rm C}L_{\gamma,\h}^{\rm C}+\varepsilon^{\rm pair}L_{\gamma,\h}^{\rm pair}\,,\ee
where $\varepsilon^{\rm C}=2.01$ and $\varepsilon^{\rm pair}=\varepsilon^{\rm BS}$; the latter is because the pair production
and BS are symmetric processes (see Appendix~\ref{bremsssigma}). The dominant contribution to $L^{\rm
PI}_{\gamma}$ is provided by the K-shell photoionisation of heavy species (see e.g. Draine~\cite{d11}). Detailed
expressions for the three terms on the right-hand side of Eq.~(\ref{lossphotons}) are given in Appendix~\ref{Lphoton}.

\begin{figure}[!ht]
\begin{center}
\resizebox{\hsize}{!}{\includegraphics{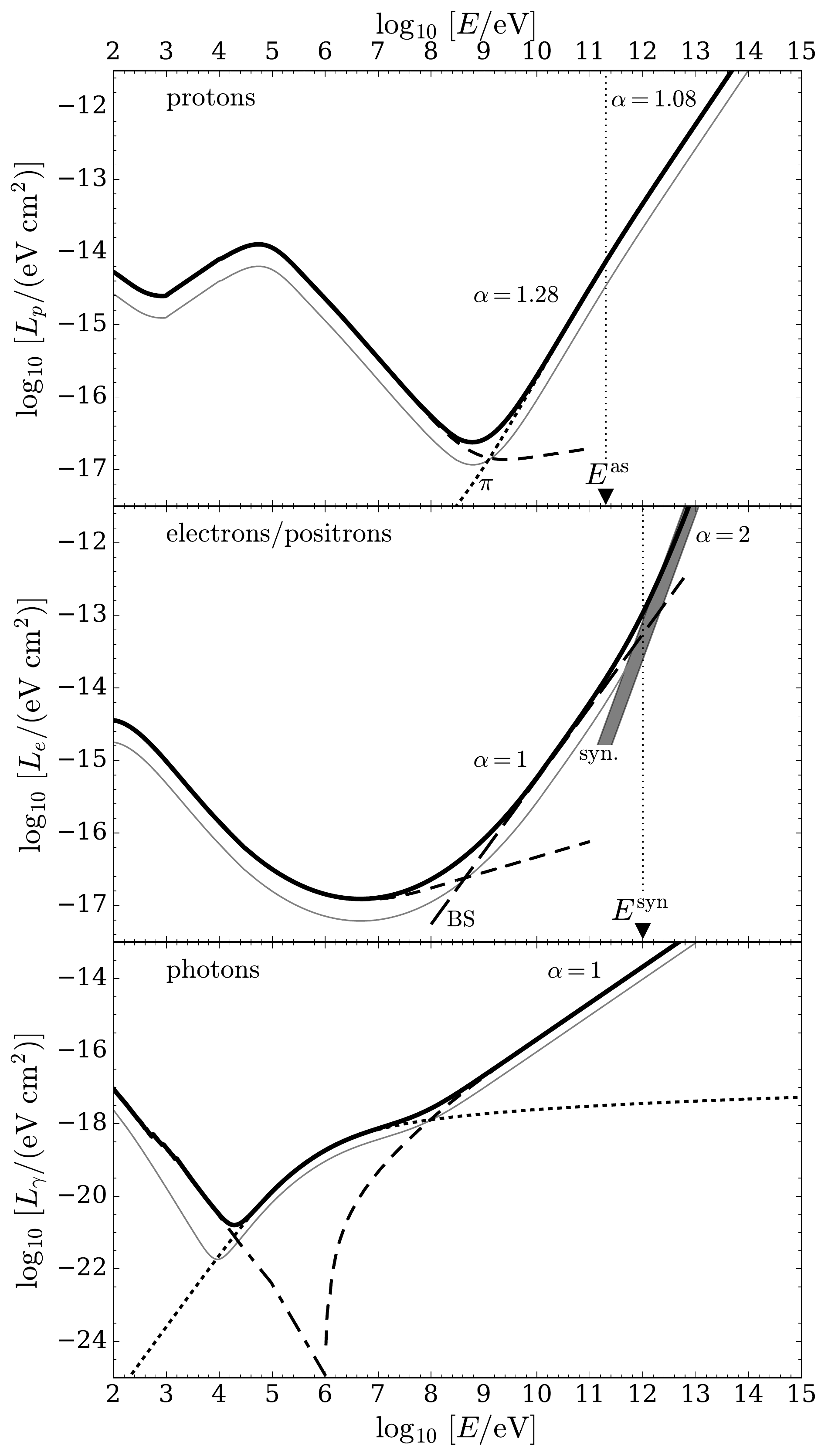}}
\caption{Total energy loss functions
of primary and secondary CR particles $k$ (protons, electrons and positrons, and photons), computed for a medium composition given in 
Table~\ref{tabwilms} ($L_k$, {\em thick black lines}) and for atomic hydrogen
($L_{k,\h}$, {\em thin grey lines}). {\it Protons} (upper panel): ionisation losses ({\em short dashed lines})  
and pion production ({\em dotted lines});
the vertical dotted line shows the energy, $E^{\rm as}$, at which 
the proton loss function changes its asymptotic behaviour from $\alpha=1.28$ to $\alpha=1.08$.
{\it Electrons and positrons} (middle panel): ionisation losses ({\em short dashed line}), BS ({\em long-dashed line}), and synchrotron losses with the uncertainty on the magnetic field strength ({\em
shaded area}, see Appendix~\ref{Lelectron});
the vertical dotted line shows the transition energy, $E^{\rm syn}$, between
BS ($\alpha=1$) and synchrotron ($\alpha=2$) losses. 
{\it Photons} (lower panel): photoionisation losses ({\em dash-dotted line}),
Compton scattering ({\em dotted line}), and pair production ({\em short-dashed line}).}
 \label{Lfunc}
\end{center}
\end{figure}

\subsection{Loss functions}

Figure~\ref{Lfunc} shows the total proton, electron, positron, and photon energy loss functions $L_k(E)$ calculated for the
IS medium composition given in Table~\ref{tabwilms}. For comparison, the loss functions in a medium of pure H atoms,
$L_{k,\h}(E)$, are also plotted. Data for the ionisation by protons are taken from the 
Stopping and Range of Ions in Matter package (Ziegler et
al.~\cite{zz10}); for the ionisation by electrons we adopt Dalgarno et al.~(\cite{dy99}) and the
National Institute of Standards and Technology database\footnote{\tt
http://physics.nist.gov/PhysRefData/Star/Text}.

We notice a change in the asymptotic behaviour $L_p\propto E^{\alpha}$ of the proton loss function, from $\alpha=1.28$
to $\alpha=1.08$, occurring at energy $E^{\rm as}$ (see Eq.~\ref{pionlosses}). 
The asymptotic
behaviour of the electron and positron loss function changes from $\alpha=1$ to $\alpha=2$, due to the transition from the BS to
synchrotron (catastrophic) losses at energy $E^{\rm syn}$ (see Appendix~\ref{Lelectron}). As for photons,
the asymptotic behaviour of the loss
function is determined by the pair production (catastrophic) losses with $\alpha=1$; at low energies, where photoionisation
dominates, one can see small peaks (around 1~keV) due to K-shell ionisation of heavy species of the medium.

\subsection{CSDA and catastrophic regimes}

The continuous slowing-down approximation (hereafter CSDA; Takayanagi 1973) has been used so far to study the propagation of
low-energy CRs in molecular clouds. It is based on two assumptions: ({\em i}\/) that the energy losses are continuous,
and ({\em ii}\/) that the pitch-angle scattering (with respect to the local magnetic field) is negligible. If these
assumptions are justified, then the loss function $L_k(E)$ (Eq.~\ref{loss_cont}) entirely determines
the modification and attenuation of the spectrum of a species $k$ with column density $N$.

\begin{figure}[!htb]
\begin{center}
\resizebox{\hsize}{!}{\includegraphics{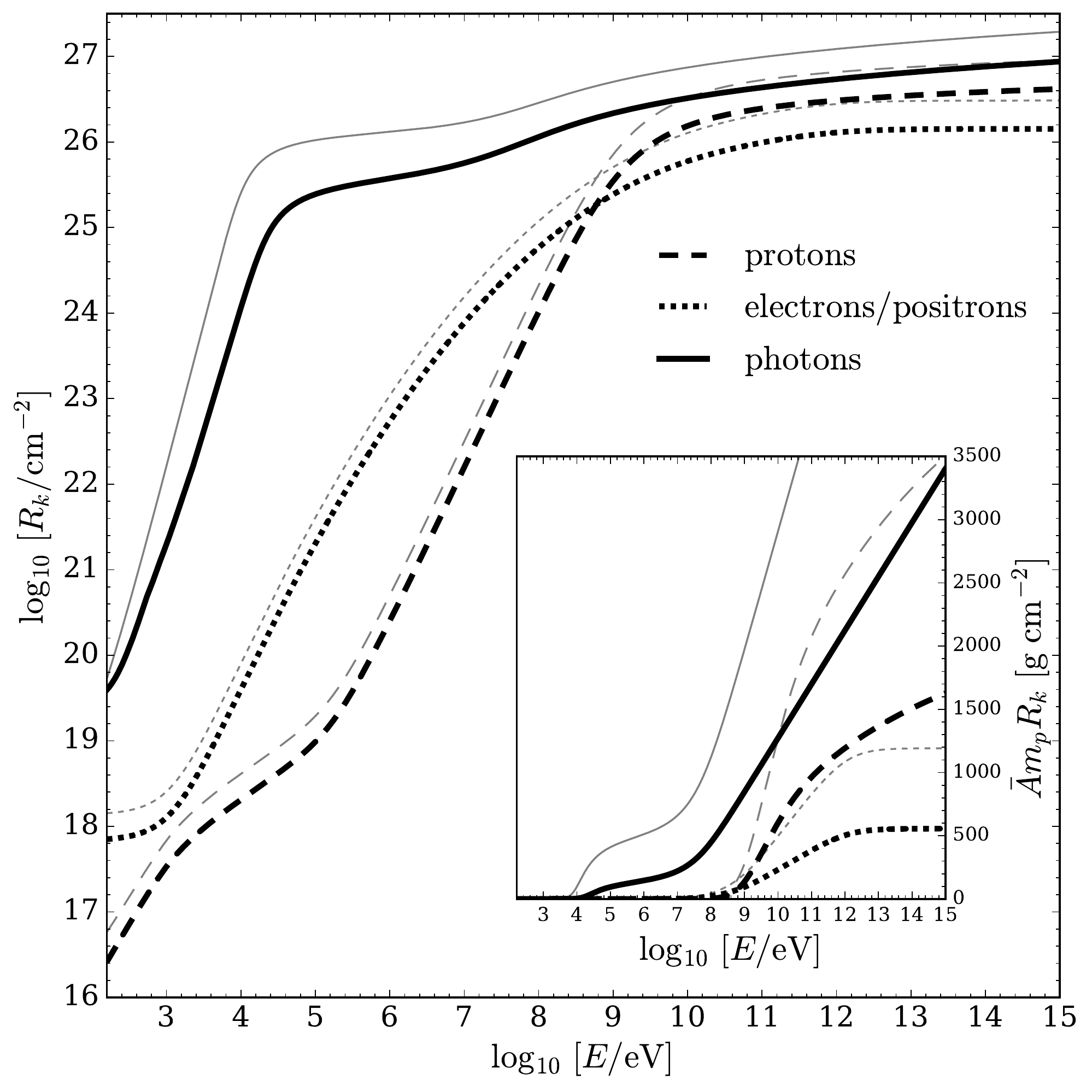}}
\caption{Total range functions, $R_k(E)$, of primary and secondary CR particles ({\em thick black lines}),  Eq.~(\ref{range_0}).
The inset shows the total range functions multiplied by $\bar A m_{p}$, to highlight the behaviour at large surface densities.
For comparison, the range functions for atomic hydrogen are also plotted ({\em thin grey lines}).} \label{fig1bis}
\end{center}
\end{figure}

Figure~\ref{fig1bis} presents the so-called range functions
\begin{equation}\label{range_0}
R_{k}(E)=\int_{0}^{E}\frac{\ud E'}{L_{k}(E^\prime)}.
\end{equation}
In the CSDA framework, the kinetic energy of a CR particle decreases from $E_0$ to $E$ after traversing a column density
\begin{equation}\label{range}
N= R_k(E_0)-R_k(E).
\end{equation}
The local spectrum at that $N$ and energy $E$ is then related to the IS spectrum at $N=0$ and energy $E_0$
via (see PGG09)
\begin{equation}\label{eqcsda}
L_k(E)j_k(E,N/\mu)=\frac12L_k(E_0)j_k^{\rm IS}(E_0),
\end{equation}
where $\mu$ is the cosine of the pitch angle. The factor of 1/2 in Eq.~(\ref{eqcsda}) takes into account that only half of
the IS CRs penetrates into the semi-infinite medium. This relation directly follows from a solution of the
transport equation for the CSDA regime (Ginzburg \& Syrovatskii~\cite{gs64}, Berezinskii et~al.~\cite{Ber90}),
\begin{equation}
\label{transport_eq}
\mu\frac{\partial j_k}{\partial N}-\frac{\partial}{\partial E}\left(L_kj_k\right)=0,
\end{equation}
assuming no sources. As pointed out in Sect.~\ref{energylosses}, it is
sufficient to analyse the CR propagation for $\mu=1$ to calculate
the total ionisation rate (see Sect.~\ref{ionisation}). The CSDA is a very simple and efficient
approach which, of course, has certain limitations (see also Sect.~\ref{propagation}).

The CSDA is formally no longer applicable when catastrophic losses dominate, although in some cases it may still be used
(with the corresponding loss function, Eq.~\ref{loss_cat}) to qualitatively describe propagation of CRs. Strictly speaking,
when both continuous and catastrophic loss processes are present, $L_k$ in Eq.~(\ref{transport_eq}) should include only the
continuous processes, while the catastrophic processes (with the cross section $\sigma_k$) should be described by an
additional term $\sigma_kj_k$ on the left-hand side. In the following, we discuss the effect of catastrophic losses on propagation of
high-energy CR electrons (Sect.~\ref{propagatione}) and employ a transport equation for this regime to describe propagation
of secondary photons (Sect.~\ref{solution_cat}).

\section{Propagation of CRs at high column densities}
\label{propagation}

\subsection{Cosmic- ray protons}
\label{propagationp}

\begin{figure}[!tb]
\begin{center}
\resizebox{\hsize}{!}{\includegraphics{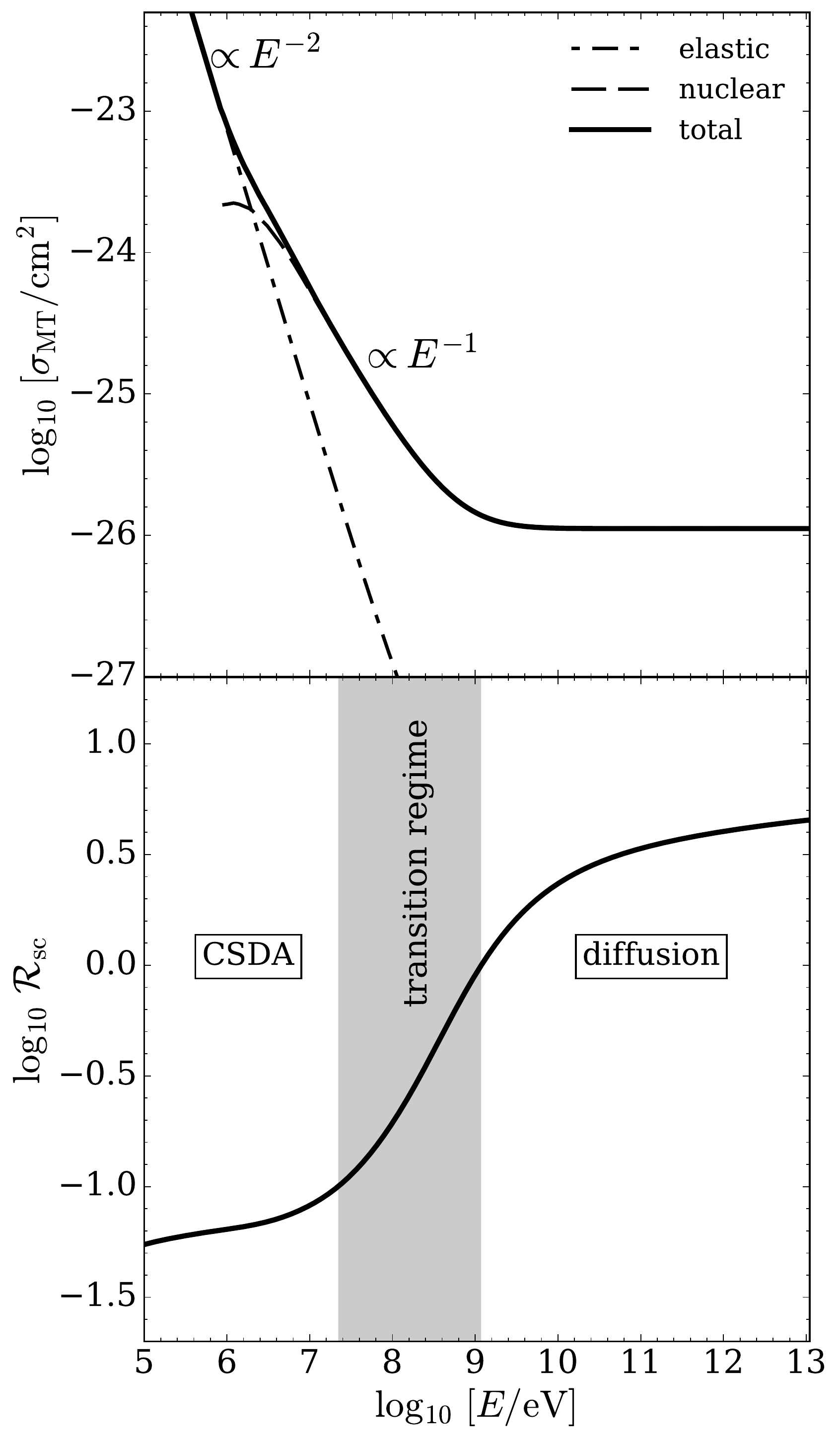}}
\caption{Upper panel: Total momentum transfer cross section for proton-nucleus collisions, $\sigma_{\rm MT}(E)$.
The elastic (Coulomb) scattering dominating at lower energies crosses over to the nuclear scattering at higher energies.
Lower panel: scattering parameter ${\cal R}_{\rm sc}(E)$, Eq.~(\ref{Rsc}). The vertical grey
stripe indicates a continuous transition from the CSDA regime, where ${\cal R}_{\rm sc}\ll1$ and the proton scattering is
unimportant, to the diffusive regime, where ${\cal R}_{\rm sc}\gtrsim1$.}
\label{sigmt}
\end{center}
\end{figure}

At energies larger than $E^{\pi}$ the interaction between CR protons and the medium leads to the production of
pions. Since the pion rest mass is significant, CR protons lose a non-negligible fraction of their energy in each collision (Schlickeiser~\cite{s02}). Such losses
cannot be formally considered as continuous, but treating them as catastrophic is not quite correct either since many
collisions are still required to reduce the energy significantly.
Furthermore, the role of elastic scattering between CR protons and target nuclei becomes increasingly important
when pion production dominates losses because the relative contribution of the nuclear interactions increases. This
effect can be understood by considering the momentum transfer cross section, $\sigma_{\rm MT}(E)$, which is made
up of various contributions depending on the energy range (see upper panel of Fig.~\ref{sigmt}). Below about 1~MeV the
elastic (Coulomb) scattering dominates, while at higher energies the CR proton interacts with the target nucleus.
Between 1~MeV and 10~MeV there is a transition region where the elastic scattering ($\sigma_{\rm MT}\propto E^{-2}$) is
modified by nuclear forces ($\sigma_{\rm MT}\propto E^{-1}$). Finally, above $\approx1$~GeV the momentum transfer cross
section becomes energy independent, which is a manifestation of the hard sphere-like scattering.\footnote{Other contributions at
lower energies are described in Jackson \& Blatt~(\cite{jb50}).} 
The total momentum transfer cross section can be written as a function of the proton-proton momentum transfer cross
section (Appendix~\ref{mtsigmaeqs}) as
\be
\sigma_{\rm MT}=\xi\sigma_{\rm MT}^{\rm pp}\,,
\ee
where $\xi=1.49$ accounts for heavy species in the ambient medium (see Eq.~\ref{xifac}).

One can easily quantify the relative importance of the elastic scattering for CR protons, as compared to the their
attenuation (characterised by loss function $L_p$). We introduce the scattering parameter,
\begin{equation}\label{Rsc}
{\cal R}_{\rm sc}(E)=\int_{0}^{E}\frac{\sigma_{\rm MT}(E^\prime)}{L_p(E^\prime)}\ud E'\,
,\end{equation}
which is the integral ratio of the characteristic stopping range owing to energy losses, $\int_0^E\ud E'/L_p$, to the
characteristic column density resulting in strong scattering, $\sim1/\sigma_{\rm MT}$. The CSDA is
a good approximation
as long as ${\cal R}_{\rm sc}(E)\ll1$, otherwise scattering is no longer negligible and a gradual transition to the
diffusive transport takes place.

The lower panel of Fig.~\ref{sigmt} shows the dependence ${\cal R}_{\rm sc}$ versus $E$. The CSDA works for
$E\lesssim25$~MeV, where ${\cal R}_{\rm sc}\ll1$; according to Fig.~\ref{fig1bis}, this corresponds to column
densities $N\lesssim7\times10^{22}$~cm$^{-2}$. The scattering of CR protons becomes increasingly important at higher
energies, and the diffusive regime operates above $\approx 1$~GeV, where ${\cal R}_{\rm sc}>1$, corresponding to
$N\gtrsim2\times10^{25}$~cm$^{-2}$.

Thus, a CSDA solution obtained for the local spectrum of CR protons at low energies should be combined with the diffusion
solution at high energies. In Appendix~\ref{match} we describe the matching procedure for the two solutions. The exact value
of the transition energy $E^{\rm tr}$ at which the solutions are matched, $25$~MeV~$\lesssim E^{\rm tr}\lesssim1$~GeV,
turns out to have a minor effect on the final results. For example, the resulting ionisation rate $\zeta(N)$ varies by less than
10\% in the corresponding range of column densities of $7\times10^{22}~{\rm cm}^{-2}\lesssim N\lesssim
2\times10^{25}$~cm$^{-2}$.

We now obtain the solution for the diffusive regime, assuming continuous losses. The steady-state density of CR protons
in the diffusive regime is governed by the following equation (Ginzburg \& Syrovatskii~\cite{gs64}),
\begin{equation}\label{ginz}
D_p\frac{\partial^{2}{\cal N}_{p}}{\partial \ell^{2}}=\frac{\partial}{\partial E}\left(\frac{{\rm d}E_p}{{\rm d}t}{\cal
N}_{p}\right)\,.
\end{equation}
Here,
$\ell$ is the coordinate along the local magnetic field (CR path),
${\cal N}_{p}(E,\ell)$ is the number of protons per unit volume and energy,
related to $j_{p}(E,\ell)$
via
\begin{equation}
{\cal N}_{p}(E,\ell)=\frac{4\pi j_{p}(E,\ell)}{\beta c}\,,
\end{equation}
and
\begin{equation}
\frac{{\rm d}E_p}{{\rm d} t}\equiv -n\beta c L_{p}(E) \,,
\label{defloss}
\end{equation}
where $\beta$ is the proton velocity in unit of the speed of light, $c$.
The diffusion coefficient is
\begin{equation}\label{Dp}
D_p(E)\approx\frac{\beta c}{3n\sigma_{\rm MT}(E)}\,,
\end{equation}
where $n$ is the total particle number density, summed over all the medium species and the factor 3 accounts for
the fact that
diffusion occurs
in three dimensions.
Then, by substituting the definition of the energy loss function for protons, Eq.~(\ref{defloss}), introducing the time-like coordinate\footnote{According to Eq.~(\ref{pionlosses}), $L_p(E)$ increases asymptotically faster than linearly, and
therefore $T(E)$ remains finite.} 
\begin{equation} \label{ginz2}
T(E)=\frac{1}{3}\int_{E}^{\infty}\frac{\ud E'}{\sigma_{\rm MT}(E^\prime)L_{p}(E^\prime)}\,,
\end{equation}
and taking into account that $\ud N/\ud \ell=n$, we reduce Eq.~(\ref{ginz}) to
\begin{equation}\label{ginz3}
\frac{\partial {\cal F}_{p}}{\partial T}=\frac{\partial^{2}{\cal F}_{p}}{\partial N^{2}}\,,
\end{equation}
where ${\cal F}_{p}=\beta L_{p}j_{p}$. Eq.~(\ref{ginz3}) must be solved with the boundary condition ${\cal
F}_{p}(T,N=0)=\beta L_{p}j_{p}^{\rm IS}/2$, since only half of the IS flux penetrates into the semi-infinite
medium, and zero initial condition, reflecting the fact that $j_{p}(E,N)\rightarrow0$ for $E\to\infty$. In analogy
with the solution of the problem of heat diffusion in a half-space (Tikhonov \& Samarskii~\cite{ts13}), the solution is
\begin{eqnarray}\label{diffsol}
j_{p}(E,N) &&= \frac{N}{12\sqrt{\pi}\beta(E) L_{p}(E)}\times\\
&&\int_{E}^{\infty}\frac{\beta(E_{0})j_p^{\rm IS}(E_{0})}{\sigma_{\rm MT}(E_{0})T^{3/2}(E,E_{0})}
\exp\left[-\frac{N^{2}}{4T(E,E_{0})}\right]\ud E_{0}\,,\nonumber
\end{eqnarray}
where $T(E,E_{0})\equiv T(E)-T(E_0)$.

\begin{figure}[]
\begin{center}
\resizebox{\hsize}{!}{\includegraphics{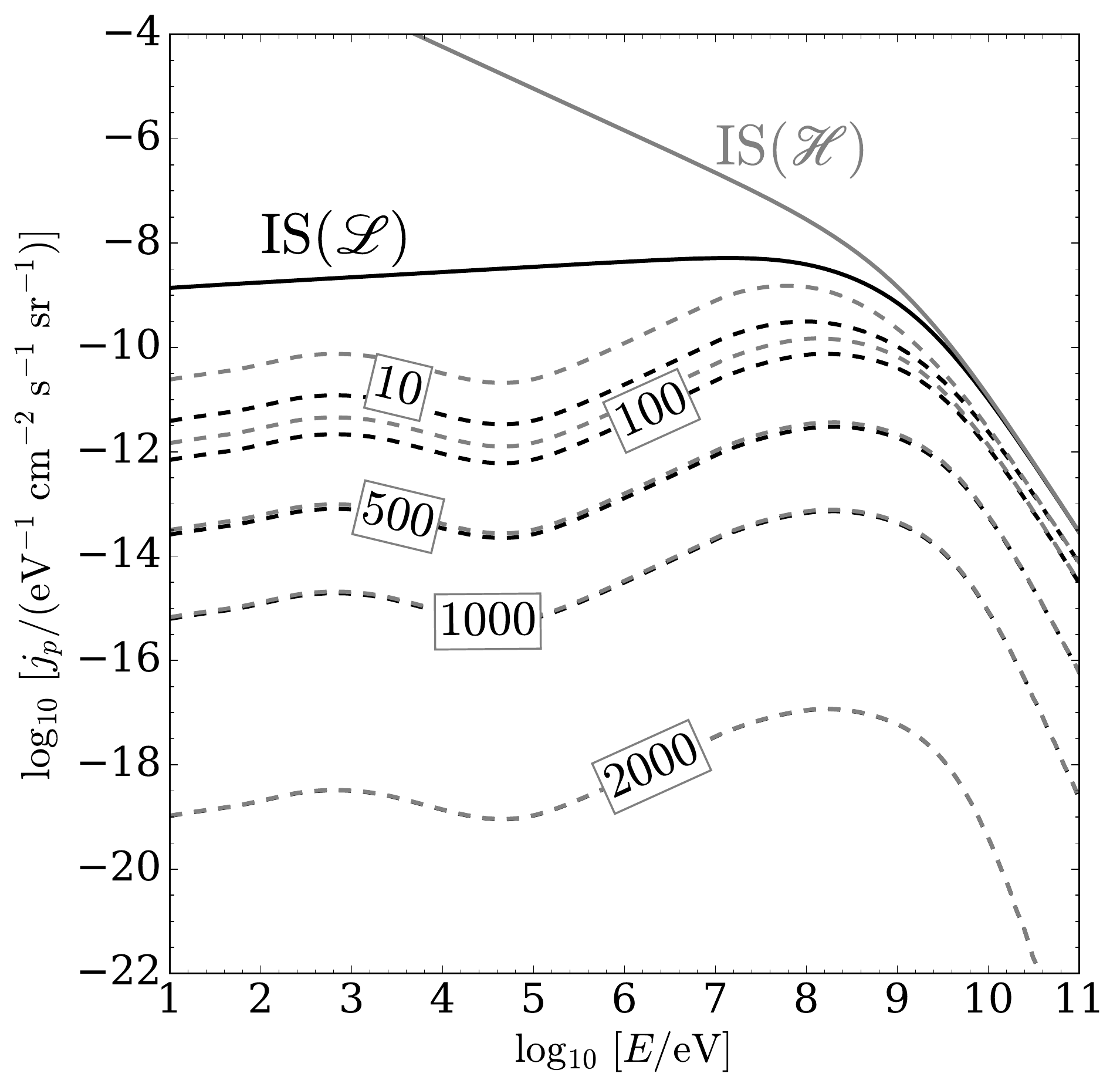}}
\caption{Interstellar (solid lines, labelled ``IS'') and local (dashed lines) differential fluxes (or spectra) of CR protons.
Only half the IS proton flux penetrates into the semi-infinite medium.
Labels indicate the surface density in g~cm$^{-2}$ for the local spectra. 
Models of the IS proton spectra,
$\mathscr{L}$ (black) and $\mathscr{H}$ (grey), are described by Eq.~(\ref{jis}).}
\label{pspectra}
\end{center}
\end{figure}

In principle, elastic scattering of CR protons leads to new source and sink terms in the general
transport equation associated with efficient energy exchange with hydrogen nuclei. In Appendix~\ref{elastic} we present
a detailed discussion of this effect, and show that for realistic conditions these new terms can be safely neglected.
Figure~\ref{pspectra} shows local differential fluxes of CR protons at various surface densities
$\Sigma$.

\subsection{Cosmic-ray electrons}
\label{propagatione}

Energy losses of electrons by BS overcome ionisation losses above $E^{\rm BS}\approx500$~MeV (see middle panel
of Fig.~\ref{Lfunc}). As pointed out by Ginzburg \& Syrovatskii~(\cite{gs64}), the energy of a photon created by
BS is generally of the order of the energy of the electron that generated it.
Therefore, one can approximately treat BS as a catastrophic process.
The effective cross section of
this process corresponds to a column density of $N^{\rm BS}\approx1.5\times10^{25}$~cm$^{-2}$, i.e.
$\approx 58$~g~cm$^{-2}$.

As a consequence, CSDA slightly overestimates the electron population at $E\gtrsim E^{\rm BS}$. However, electrons at these
energies yield only a minor contribution to the ionisation rate. Our numerical results show that their effect is
smaller than 2\% at $\Sigma\approx 30$~g~cm$^{-2}$, and becomes vanishingly small at higher $\Sigma$. Furthermore, the
results presented below in Sect.~\ref{ionisation} demonstrate that -- even in the CSDA regime -- the ionisation by
primary CR electrons is negligible compared to the contribution from CR protons at $\Sigma>1$~g~cm$^{-2}$ for all models of
IS CRs considered in this paper.

\section{Generation and propagation of secondary particles}
\label{generation}

\begin{figure}[!htb]
\begin{center}
\resizebox{\hsize}{!}{\includegraphics{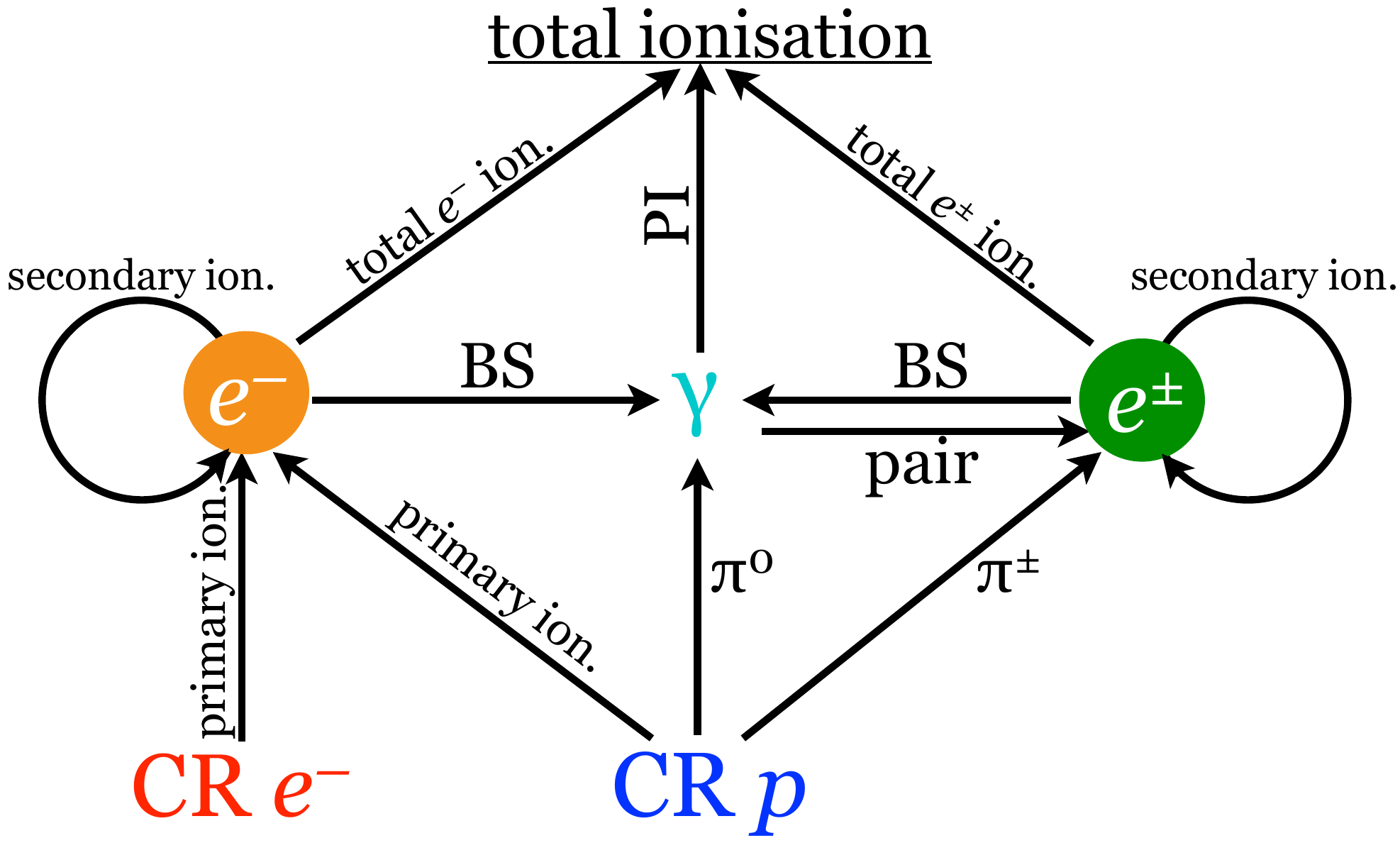}}
\caption{Ionisation diagram, explaining the effect of secondary particles that are generated (directly or indirectly) by
CR protons and electrons through ionisation, pion decay ($\pi^{0},\pi^{\pm}$),  BS, and pair production
(pair). The secondary particles include electrons ($e^-$, due to primary and secondary ionisation), positrons and electrons
($e^{\pm}$, due to pair production and $\pi^{\pm}$ decay; also electrons produced in secondary ionisation are
included), and photons ($\gamma$, due to BS and $\pi^0$ decay), all contributing to the respective ionisation routes.}
\label{diagram}
\end{center}
\end{figure}

Figure~\ref{diagram} presents the main ionisation routes associated with various secondary particles that can be produced
by CRs in circumstellar discs. Cosmic-ray protons and electrons generate secondary electrons by primary ionisation ($p_{\rm CR}+{\rm
H_{2}}\rightarrow p_{\rm CR}'+{\rm H_{2}^{+}}+e^{-}$). Secondary electrons with energy larger than the H$_{2}$ ionisation
potential produce further generations of ambient electrons. Cosmic-ray protons colliding with  protons create charged pions;
through muon decay, they become electrons and positrons ($\pi^{\pm}\rightarrow\mu^{\pm}\rightarrow e^{\pm}$), producing
secondary ionisations. In addition, proton-proton collisions create neutral pions decaying into photon pairs
($\pi^{0}\rightarrow2\gamma$). The second important source of photons is BS by electrons and positrons. One
should also account for electron-positron pair production by photons ($\gamma\rightarrow e^++e^-$). In the following we give
the equations needed to compute the differential flux of all the secondary particles.

\subsection{Photons}
\label{photons_sec}

\begin{figure}[!htb]
\begin{center}
\resizebox{\hsize}{!}{\includegraphics{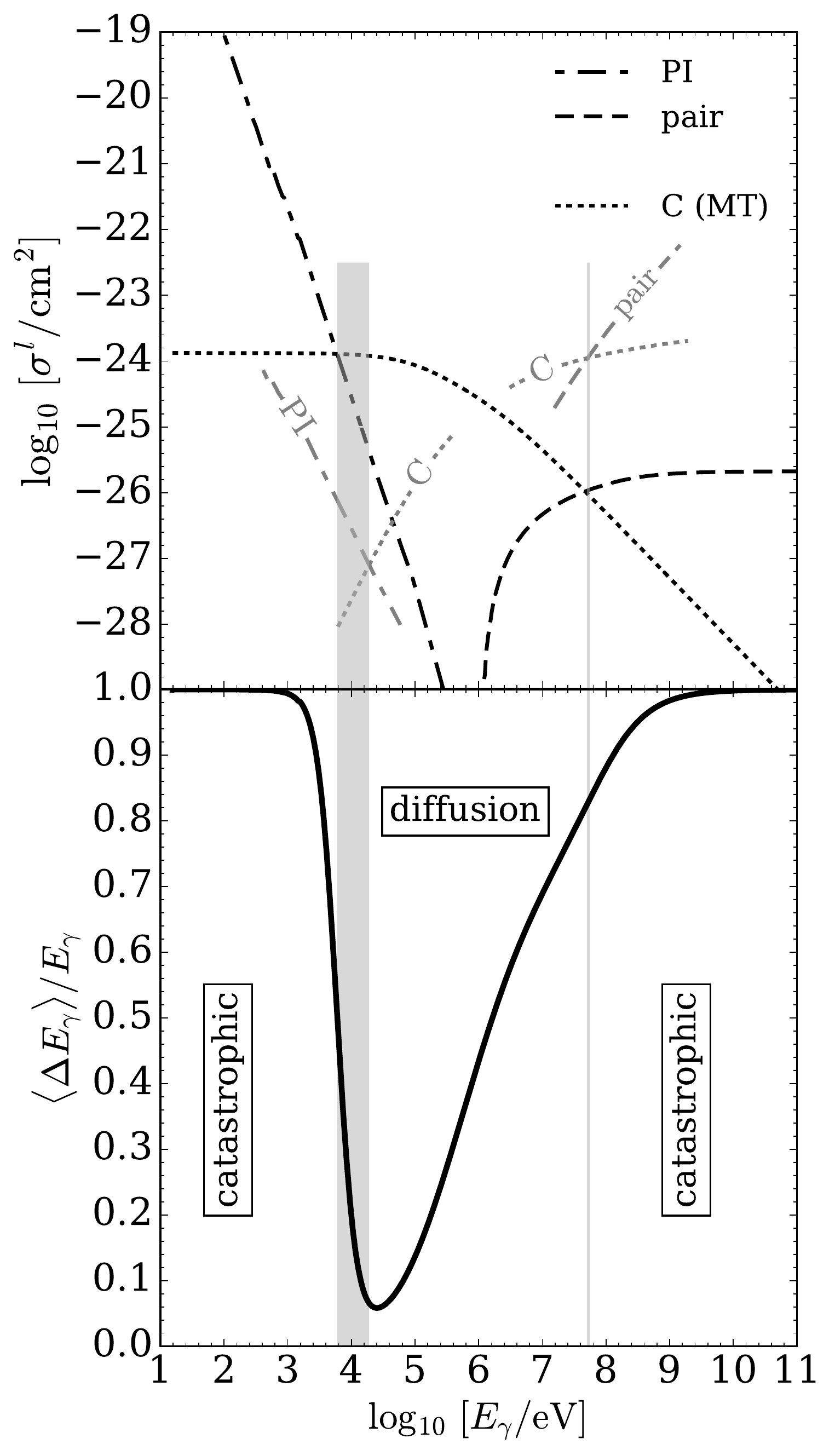}}
\caption{Upper panel:
Components of the cross section for photons interacting with nuclei via process $l$: photoionisation ($\sigma^{\rm PI}$), Compton
scattering ($\sigma^{\rm C}$), and pair production ($\sigma^{\rm pair}$). The momentum-transfer (MT) cross section
is plotted for Compton scattering, while for catastrophic (PI and pair) processes the cross section coincides with the
MT cross section. The grey lines depict the corresponding loss functions near to their crossing (in arbitrary units). The
vertical grey stripes, indicating the energy intervals between the crossing points of the corresponding cross sections and
loss functions, separate the diffusive and catastrophic regimes of the photon transport (see text for details). Lower panel:
the relative average energy lost by a photon per collision, $\langle\Delta E_\gamma \rangle/E_{\gamma}$, vs. the photon
energy, Eq.~(\ref{Delta_E}).} \label{fig4a}
\end{center}
\end{figure}

The regimes of propagation for photons can be different depending on their energy. As shown in the upper panel of
Fig.~\ref{fig4a}, photoionisation and pair production dominate below $\approx 5$~keV and above $\approx 50$~MeV, respectively.
Both processes are catastrophic, i.e. photons disappear after the interaction with nuclei.
As for Compton scattering,
the relative average energy lost by a photon in each interaction with electrons is written\begin{eqnarray}
\left.\frac{\langle\Delta E\rangle}{E}\right|^{\rm C}=\frac{1}{E\sigma^{\rm C}(E)}
\int_{0}^{E_{e}^{\rm max}}E_{e}\frac{\ud\sigma^{\rm C}(E,E_e)}{\ud E_{e}}\ud E_{e}
=\frac{L_{\gamma}^{\rm C}(E)}{E\sigma^{\rm C}(E)}\,,
\end{eqnarray}
where $\sigma^{\rm C}(E)$ is the Compton cross section (see Eq.~\ref{sigCtot}). For $x=E/(m_{e}c^{2})\gg1$ we have
$\langle\Delta E\rangle/E|^{\rm C}\approx  1-4/(3\ln x)\to1$, i.e. Compton losses become catastrophic. On the other hand, for
$x\ll1$ photons transfer a small fraction of their energy, $\langle\Delta E\rangle/E|^{\rm C}\approx 3x/2$, and losses are
continuous. However, below $E\approx1$~keV photoionisation is the dominant process, and losses become catastrophic again.
This is shown in the lower panel of Fig.~\ref{fig4a}, where we plot the quantity $\langle\Delta E\rangle/E|_{\gamma}$
combining all energy loss processes for photons,
\begin{equation}\label{Delta_E}
\left.\frac{\langle\Delta E\rangle}{E}\right|_{\gamma}=\frac{\sigma^{\rm PI}+L_{\gamma}^{\rm C}/E+\sigma^{\rm pair}}
{\sigma^{\rm PI}+\sigma^{\rm C}+\sigma^{\rm pair}}\,.
\end{equation}
In order to determine whether CSDA or diffusive regime is applicable in the intermediate energy range of $5~{\rm
keV}\lesssim E\lesssim 50~{\rm MeV}$, we compute the scattering parameter ${\cal R}_{\rm sc}(E)$ defined by Eq.~(\ref{Rsc}).
We integrate the ratio of the momentum transfer cross section $\sigma_{\rm MT}^{\rm C}(E)$ (Eq.~\ref{sigmamtC}) and the loss
function for Compton scattering $L_{\gamma}^{\rm C}(E)$ (see Sect.~\ref{Lphoton}), which yields ${\cal R}_{\rm
sc}\approx3\times10^{4}$ in this energy range. This clearly implies a diffusive regime for photons with dominant Compton
scattering.

In the following subsections we present the governing equations for the catastrophic and diffusive regimes, and discuss how
their solutions can be matched.

\subsubsection{Catastrophic regime}
\label{solution_cat}

The equation of radiative transfer for the differential flux of photons $j_\gamma(E,N)$ (photon flux per unit energy and
solid angle) is
\begin{equation}\label{radtr}
\mu\frac{\partial j_\gamma(E,N)}{\partial N}+\sigma_\gamma(E)j_\gamma(E,N)=S_{\gamma}(E,N).
\end{equation}
Here, $\mu$ is the cosine of the angle with respect to the direction of CR propagation and 
$\sigma_\gamma=\sigma^{\rm
PI}+\sigma^{\rm pair}$ is the cross section accounting for the two catastrophic processes described in the previous section, i.e.
photoionisation and pair production (see Appendices~\ref{phionsigma} and \ref{bremsssigma}). The source function of
photons, $S_\gamma(E,N)$, namely the number of photons per unit time, energy, and solid angle produced per  nucleus,
\begin{equation}\label{Stot}
S_\gamma=S^{\pi^{0}}+S^{\rm BS}
,\end{equation}
is the sum of the source function for $\pi^{0}$ decay from proton-nucleus collisions,
\begin{equation}\label{Spp}
S^{\pi^{0}}(E,N)=2\varepsilon^{\pi}\int_{E+E^{\pi}}^\infty j_p(E_p,N)\frac{\ud\sigma^{\pi^{0}}_{\h}(E_p, E)}{\ud E}\, \ud E_p,
\end{equation}
(each proton provides 2 photons), and the source function for BS
\begin{equation}\label{SBS}
S^{\rm BS}(E,N)=2\varepsilon^{\rm BS}\int_{E}^\infty j_e(E_e,N)\frac{\ud \sigma^{\rm BS}_{\h}(E_e, E)}{\ud E}\, \ud E_e\,,
\end{equation}
where the factor 2 accounts for electrons and positrons. Here, $\ud\sigma_{\h}^{\pi^{0}}/\ud E$ is the differential
cross section for photon production by $\pi^0$ decay (Kamae et al.~\cite{kk06}), and $\ud\sigma_{\h}^{\rm BS}/\ud E$ is
the differential cross section of atomic hydrogen for BS (Blumenthal \& Gould~1970; see
Eq.~\ref{BHbremss}).

Neglecting any source of photon radiation external to the cloud (i.e. at $N=0$), and averaging over $\mu$, the solution of
Eq.~(\ref{radtr}) gives the photon differential flux $j_{\gamma}^{\rm cat}$ in the catastrophic regime,
\be\label{jgammacat} j_{\gamma}^{\rm cat}(E,N) \!=\! \frac12\int_0^\infty S_\gamma(E,N')\,\ud N' \!\!\int_{0}^{1}
\exp\left[-\frac{\sigma_\gamma(E)|N-N'|}{\mu}\right] \, \frac{\ud\mu}{\mu}\,. \ee
The factor 1/2 in Eq.~(\ref{jgammacat}) takes into account the fact that only the
forward (backward) propagating photons produced at $N^{\prime}<N$ ($N^{\prime}>N$) contribute to the local differential
flux, $j_{\gamma}^{\rm cat}$ at a given column density $N$.

\subsubsection{Diffusive regime}
\label{solution_diff}

The diffusive regime of photons is conceptually similar to that of CR protons and therefore is described by a similar equation
(Eq.~\ref{ginz}), but with additional source terms owing to the photon production by neutral pion decay and BS
(Eqs.~\ref{Spp} and~\ref{SBS}, respectively). The diffusion equation is then given by
\begin{equation}\label{eqphot}
D_{\gamma}\frac{\partial^{2}{\cal N}_{\gamma}}{\partial \ell^{2}}=\frac{\partial}{\partial E}
\left(\frac{{\rm d}E_{\gamma}}{{\rm d}t}{\cal N}_{\gamma}\right)+4\pi n S_{\gamma}(E,N),
\end{equation}
where the number of photons per unit volume and energy is ${\cal N}_{\gamma}(E,N)=4\pi j_{\gamma}(E,N)/c$. Substituting
${\rm d}E_{\gamma}/{\rm d}t=-ncL_{\gamma}^{\rm C}(E)$ and
\begin{equation}
D_{\gamma}(E)\approx\frac{c}{3n\sigma_{\rm MT}^{\rm C}(E)},
\end{equation}
we reduce Eq.~(\ref{eqphot}) to
\begin{equation}\label{eqphot1}
\frac{\partial{\cal F}_{\gamma}}{\partial T}=\frac{\partial^{2}{\cal F}_{\gamma}}{\partial N^{2}}
-3\sigma_{\rm MT}^{\rm C}L_{\gamma}^{\rm C}S_\gamma\,,
\end{equation}
where ${\cal F}_{\gamma}=L_{\gamma}^{\rm C}j_{\gamma}$ is a function of column density $N$ and 
time-like coordinate
\begin{equation}\label{dT}
T(E) = \frac{1}{3}\int_{E}^{E^{\rm tr}}\frac{\ud E'}{\sigma_{\rm MT}^{\rm C}(E^\prime)L_{\gamma}^{\rm C}(E^\prime)}\,.
\end{equation}
Similar to the catastrophic regime, we can set zero boundary condition, ${\cal F}_{\gamma}(E,0)=0$, while the initial condition is ${\cal F}_{\gamma}(0,N)={\cal F}_{\gamma}^{\rm cat}(E^{\rm tr},N)$. The latter condition is determined by
matching the diffusive and catastrophic regimes at $E= E^{\rm tr}$ (specified below).

Using again the analogy with the non-homogeneous heat diffusion problem in a half-space (Tikhonov \& Samarskii~\cite{ts13}),
we obtain the following solution for the photon differential flux:
\begin{equation}\label{photonflux}
j_{\gamma}(E,N)=\frac{\mathcal{F}_{1}(E,N)+\mathcal{F}_{2}(E,N)}{L_{\gamma}^{\rm C}(E)}\,,
\end{equation}
where
\begin{eqnarray}
\mathcal{F}_{1}(E,N)&=&L_{\gamma}^{\rm C}(E^{\rm tr})\int_{0}^{\infty}j_{\gamma}^{\rm cat}(E^{\rm tr},N')
G[N,N',T(E)]\ud N',\\
\mathcal{F}_{2}(E,N)&=&\int_{0}^{\infty}\int_{E}^{E^{\rm tr}}S_{\gamma}(E',N')\\\nonumber
&&\times G[N,N',T(E)-T(E')]\ud E'\ud N',
\end{eqnarray}
are determined by the Green's function
\begin{eqnarray}
G(N,N',x)&=&\frac{1}{2\sqrt{\pi x}}\left\{\exp\left[-\frac{(N-N')^{2}}{4x}\right]\right.\\\nonumber
&&\left.-\exp\left[-\frac{(N+N')^{2}}{4x}\right]\right\}\,.
\end{eqnarray}

In Sect.~\ref{photons_sec} we mentioned that the catastrophic solution obtained in Sect.~\ref{solution_cat} for the high-
and low-energy catastrophic regimes must be combined with Eq.~(\ref{photonflux}). Similar to the treatment of CR protons
(Sect.~\ref{propagationp}), we introduce the transition energy $E^{\rm tr}$ at which the two regimes should be matched. The
matching criteria are determined ({\em i}\/) by the applicability of the diffusion approximation, which requires that
$\sigma_{\rm MT}^{\rm C}j_{\gamma}\gg|\partial j_{\gamma}/\partial N|$, and ({\em ii}\/) by a continuous transition between
the solutions.

By comparing Eqs.~(\ref{radtr}) and (\ref{eqphot}) we infer that, depending on the relative importance of different terms in
the equations, $E^{\rm tr}$ may vary: as shown in Fig.~\ref{fig4a}, the matching occurs in the energy interval limited by
the intersection points of the respective cross sections or loss functions. The lower panel of Fig.~\ref{fig4a} shows that
the transition to the high-energy (pair) catastrophic regime occurs at $E^{\rm tr}\approx 50$~MeV, where both $\sigma^{\rm
C}_{\rm MT}(E)$ and $L_\gamma^{\rm C}(E)$ intersect with $\sigma^{\rm pair}(E)$ and $L_\gamma^{\rm pair}(E)$, respectively.
Concerning the matching with the low-energy (PI) catastrophic regime, the intersection of $\sigma^{\rm C}_{\rm MT}(E)$
and $\sigma^{\rm PI}(E)$ is at $E\approx 6$~keV, whereas $L_\gamma^{\rm C}(E)$ and $L_\gamma^{\rm pair}(E)$ cross at
$E\approx 20$~keV. In this case, the crossing points do not coincide, which is easy to understand: for process $k$, the loss
function is generally related to the corresponding cross section via $L^k\sim\langle\Delta E\rangle\sigma^k$. At such
energies, $\langle\Delta E\rangle/E|_{\gamma}\sim10^{-1}$ for Compton scattering, while for photoionisation
$\langle\Delta E\rangle/E|^{\rm PI}=1$ by definition. Hence, the low-energy catastrophic and diffusion solutions should be
matched at $6$~keV~$\lesssim E^{\rm tr}\lesssim20$~keV. The exact value of $E^{\rm tr}$ turns out to have a negligible
effect on the resulting ionisation rate, as for CR protons.

\subsection{Electrons and positrons}

\begin{figure*}[!htb]
\begin{center}
\resizebox{\hsize}{!}{\includegraphics{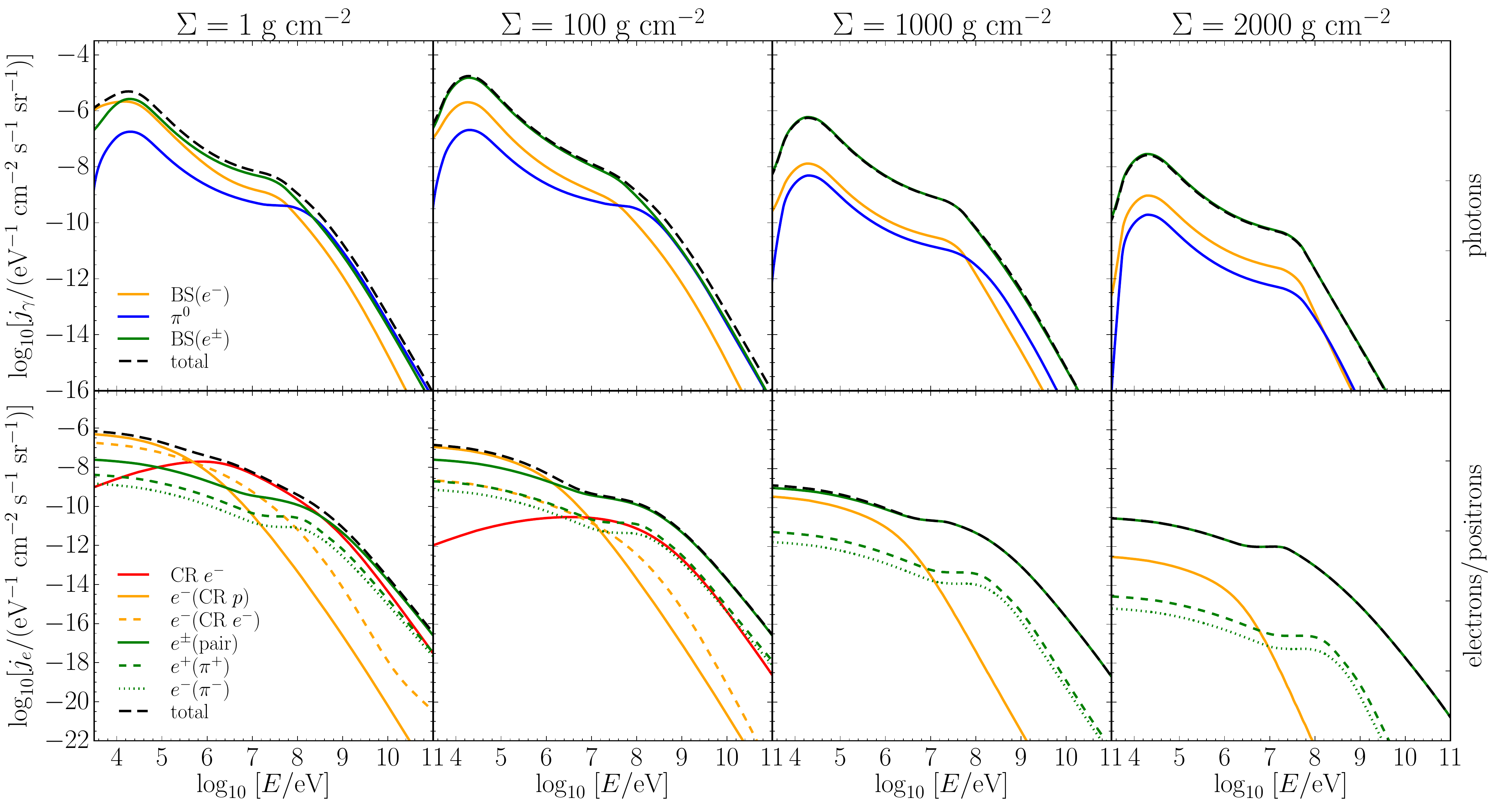}} \caption{Differential fluxes (energy spectra) $j_k(E)$ of photons (upper
row) and electrons and positrons (lower row), plotted for four characteristic values of the surface density $\Sigma$. Each plot
shows partial contributions ({\em coloured lines}) to the total differential flux ({\em black dashed line}). For photons, the
contributions are due to BS from electrons ($e^-$) produced by CRs ({\em orange line}) and due to
$\pi^{0}$ decay ({\em blue line}) and BS from electrons and positrons ($e^\pm$) created by pair production and
charged pion decay ({\em green line}). For electrons, the contributions are due to CR electrons ({\em solid red line}),
secondary electrons ($e^-$) produced by CRs ({\em solid and dashed orange lines}), and 
electrons and positrons ($e^\pm$)
generated by pair production ({\em green solid line}) and decay of charged pions ({\em dashed and dotted green lines}). }
\label{photons}
\end{center}
\end{figure*}

Electrons and positrons have two different sources. First, pairs are produced by photons with energy above
$2m_ec^2$, so that the electron and positron energy is $0 \le E \le E_\gamma-2m_ec^2$. The resulting source function at a given
column density for a single species (electron or positron),
\begin{equation}\label{Spair}
S^{\rm pair}(E,N)=\varepsilon^{\rm pair}\int_{E+2m_ec^2}^\infty j_\gamma(E_\gamma,N)
\frac{\ud\sigma^{\rm pair}_{\h}(E_\gamma,E)}{\ud E}\, \ud E_\gamma,
\end{equation}
is determined by the differential cross section $\ud\sigma_{\h}^{\rm pair}/\ud E$ (Eq.~\ref{d_sigma_pair}). Second,
electrons and positrons are created through decay of charged pions, generated in proton-nucleus collisions at energies
above $E^{\pi}$. The corresponding source function is given by
\begin{equation}\label{Spppm}
S^{\pi^{\pm}}(E,N)=\varepsilon^{\pi}\int_{E+E^{\pi}}j_{p}(E_{p},N)\frac{\ud\sigma^{\pi^{\pm}}_{\h}(E_{p},E)}{\ud E}%
\ud E_{p}\,,
\end{equation}
where $\ud\sigma^{\pi^{\pm}}_{\h}/\ud E$ is the differential cross section for electron and positron production by $\pi^{\pm}$
decay (Kamae et al.~\cite{kk06}), which we assume to have the same scaling for target heavy nuclei as that for
$\pi^{0}$ production (see Appendix~\ref{Lproton}). The total source function for 
electrons and positrons is then
\begin{equation}
S_{e^{\pm}}=S^{\rm pair}+S^{\pi^{\pm}}.
\end{equation}

As we pointed out in Sect.~\ref{propagatione}, the use of CSDA to describe propagation of
electrons and positrons
with energies above the BS threshold $E^{\rm BS}\approx 500$~MeV leads to a slight overestimation of their
differential flux. However, since the contribution of
$e^{\pm}$ of such energies to the ionisation rate is practically negligible,
CSDA can be
employed. Generally, when the stopping range is comparable to (or larger than) the local column
density $N$, the resulting spectrum of electrons and positrons is given by the convolution of the source function,
\begin{equation}\label{convolution}
j_{e^{\pm}}(E,N)=\frac{1}{2L_{e}(E)}\int_N^\infty S_{e^{\pm}}(E_0,N_0)L_{e}(E_0)\,\ud N_0,
\end{equation}
where the factor $1/2$ accounts for electrons and positrons propagating in two directions
and the initial energy $E_0>E$ at $N_0$ is related to $N$ by
\begin{equation}
|N_0-N|=\int_{E}^{E_0}\frac{\ud E'}{L_{e}(E^\prime)}\,.
\end{equation}
If the range is small, $|N_{0}-N|\ll N$, the spectrum is localised,
\begin{equation}\label{loc}
j_{e^{\pm}}(E,N)=\frac{E}{L_{e}(E)} S_{e^{\pm}}(E,N).
\end{equation}
A check a posteriori of the energy spectra calculated with Eq.~(\ref{convolution}) for $N\gtrsim10^{24}$~cm$^{-2}$ shows
that they are accurately reproduced by Eq.~(\ref{loc}).

\subsection{Differential fluxes of secondaries}

Following the diagram sketched in Fig.~\ref{diagram}, we use the following algorithm to compute the differential fluxes of
secondary CR species: In the first step, we obtain the photon flux produced by neutral pion decay and
BS, the latter being generated by electrons due to primary ionisation and by electrons and positrons from charged
pion decay. Next, we calculate the contribution to the electron and positron flux due to pair production by photons. Then,
we employ an iterative procedure for photons, electrons, and positrons until convergence.
In Fig.~\ref{photons} we present the photon, electron, and positron differential fluxes (spectra) computed for typical disc
(line-of-sight) surface densities:

{\it Photon spectrum:} At relatively low densities, $\Sigma\lesssim1$~g~cm$^{-2}$, the low-energy part of the spectrum, below
$\approx0.1$~MeV, is dominated by BS of secondary electrons created in primary ionisation
BS($e^{-}$); at energies in the range $\approx 0.1$--100~MeV, additional BS due to electrons and positrons
created by charged pion decay and pair production, BS($e^{\pm}$), becomes important; and above $\approx 100$~MeV, the spectrum
is mostly determined by neutral pion decay, $\pi^0$. When $\Sigma$ exceeds $\approx100$~g~cm$^{-2}$, the spectrum is
completely due to BS($e^{\pm}$).

{\it Electron and positron spectrum:}
At surface densities $\approx1$~g~cm$^{-2}$, the spectrum below $\approx100$~keV is dominated by secondary electrons due to
ionisation by CR protons, then CR electrons dominate up to $\approx10$~GeV, and for higher energies the contribution of
electron-positron pairs becomes the most abundant component. Above $\approx 100$~g~cm$^{-2}$, the contribution of CR
electrons becomes rapidly negligible; the spectrum below $\approx1$~MeV is dominated by secondary electrons produced by CR
protons and by electron-positron pairs at higher energies. For $\Sigma\gtrsim 1000$~g~cm$^{-2}$, the spectrum is entirely
made of electron-positron pairs. This latter fact, along with the dominance of BS($e^{\pm}$) in the photon spectrum, has a
decisive influence on the behaviour of the ionisation rate at large $\Sigma$, as discussed in Sect.~\ref{ionisation}.

\section{Ionisation at high column densities}
\label{ionisation}

The total ionisation rate of molecular hydrogen due to primary and secondary CR species $k$ (primary
protons and heavier nuclei, primary electrons, electron-positron pairs, and photons) is
\begin{equation}\label{crionint}
\zeta_{k}(N) = 4\pi\int_{I_{\rm H_{2}}}^{\infty} j_{k}(E,N)[1+\Phi_k(E)]\sigma_k^{\rm ion}(E)\,\ud E\,,
\end{equation}
where $\sigma_k^{\rm ion}(E)$ is the ionisation cross section of H$_2$  by species $k$ and $I_{\hh}=15.44$~eV is the
ionisation potential of $\hh$. For protons we adopt the ionisation cross sections by Rudd et al.~(\cite{r88}) and
Krause et al.~(\cite{km15}), for electrons we use results by Kim et al.~(\cite{ks00}), while for positrons we combine the
non-relativistic cross section by Knudsen et al.~(\cite{kb90}) with the relativistic expression for electrons. 
The effect of ionisation by secondary electrons produced by species $k$ is described by a multiplicity factor,
$\Phi_k(E)$, which is the average number of such ionisation events,
\begin{equation}
\Phi_k(E)=\frac{L_k^{\rm ion}(E)}{\langle E_e^{\rm ion} \rangle \sigma_k^{\rm ion}(E)}.
\end{equation}
Here, $\langle E_e^{\rm ion} \rangle\approx 37$~eV is the average energy lost by an electron per ionisation event (Glassgold et al.~\cite{ggp12})
and
$L_k^{\rm ion}(E)$ is the energy loss function for species $k$ due to ionisation by H$_2$. Since $L_k^{\rm ion}(E)\propto
\sigma_k^{\rm ion}(E)$ in a broad energy range (see e.g. Appendix~\ref{ion_low} for protons), the multiplicity factor
$\Phi_k(E)$ can be practically considered as a scale factor.

The contribution from charged CR species in Eq.~(\ref{crionint}) is almost entirely dominated by energies below
$\sim1$~GeV, assuming unmodulated spectra $\mathscr{L}$ and $\mathscr{H}$; the effect of CR modulation is studied in
Sect.~\ref{T-Tauri}. The propagation of protons (as well as electrons and positrons) at these energies is described by CSDA. So far we
set the pitch angle for such particles equal to zero ($\mu=1$, i.e. their velocities were assumed to be
parallel to the local magnetic field), but in fact their local spectra should be averaged over $\mu$. By performing the
averaging,
\begin{equation}\label{average}
\langle j_k(E,N)\rangle=\int_0^1 j_k\left(E,\frac{N}{\mu}\right)\,\ud \mu\equiv
N\int_N^\infty \frac{j_k(E,N')}{N'^2}\,\ud N',
\end{equation}
we notice that $\langle j_k(E,N)\rangle$ can be computed from the spectra for $\mu=1$.

The averaging over pitch angle is unimportant for electrons and positrons generated through the pair production and
$\pi^\pm$ decay: their contribution to $\zeta$ turns out to be negligible for $N\lesssim 10^{25}$~cm$^{-2}$, where the
direct ionisation by CR protons dominates (see below), while at higher $N$ their spectra are well localised (see
Eq.~\ref{loc}, i.e. their pitch angles play no role). Ionisation by CR electrons is also negligible, as we
already pointed out in Sect.~\ref{propagatione}. Thus, we need to perform the averaging only for CR protons, i.e. Eq.~(\ref{crionint}) for these values should be computed with $\langle j_p\rangle$.\footnote{The effect of the averaging
depends on the column density. For moderate values of $N$, the proton ionisation rate
can be approximated by a power-law dependence, $\zeta_p(N)\propto N^{-q}$ (see Appendix~\ref{ion_low} and
Fig.~\ref{zetavsN} in Appendix~\ref{polyfit}); then Eq.~(\ref{average}) yields 
$\langle\zeta_p(N)\rangle
=\zeta_p(N)/(1+q)$. For higher column densities, the attenuation is (roughly) exponential,
$\zeta_p(N)\propto\exp(-N/N_{\rm c})$,
with the characteristic scale $N_{\rm c}$; then $\langle \zeta_p(N)\rangle\approx \zeta_p(N)/(1+N/N_{\rm c})$.}

Summing up the contribution of CR species yields the total production rate $\zeta=\sum_{k}\zeta_{k}$ of molecular
hydrogen ions, H$_2^+$. When performing the summation, we take into account the effect of heavy CR nuclei. Since the
ionisation cross section scales as $Z^{2}$ (see PGG09) and the pion production cross section as $A^{0.79}$ (see
Geist~\cite{g91}), the proton ionisation rate is increased by a factor of 1.48 and the pair and photon ionisation rate by a
factor of 1.30;  this is the case assuming that 
heavy nuclei have the same attenuation as protons and that
CRs have the same composition as the IS medium (see Table~\ref{tabwilms}).

\begin{figure}[!htb]
\begin{center}
\resizebox{\hsize}{!}{\includegraphics{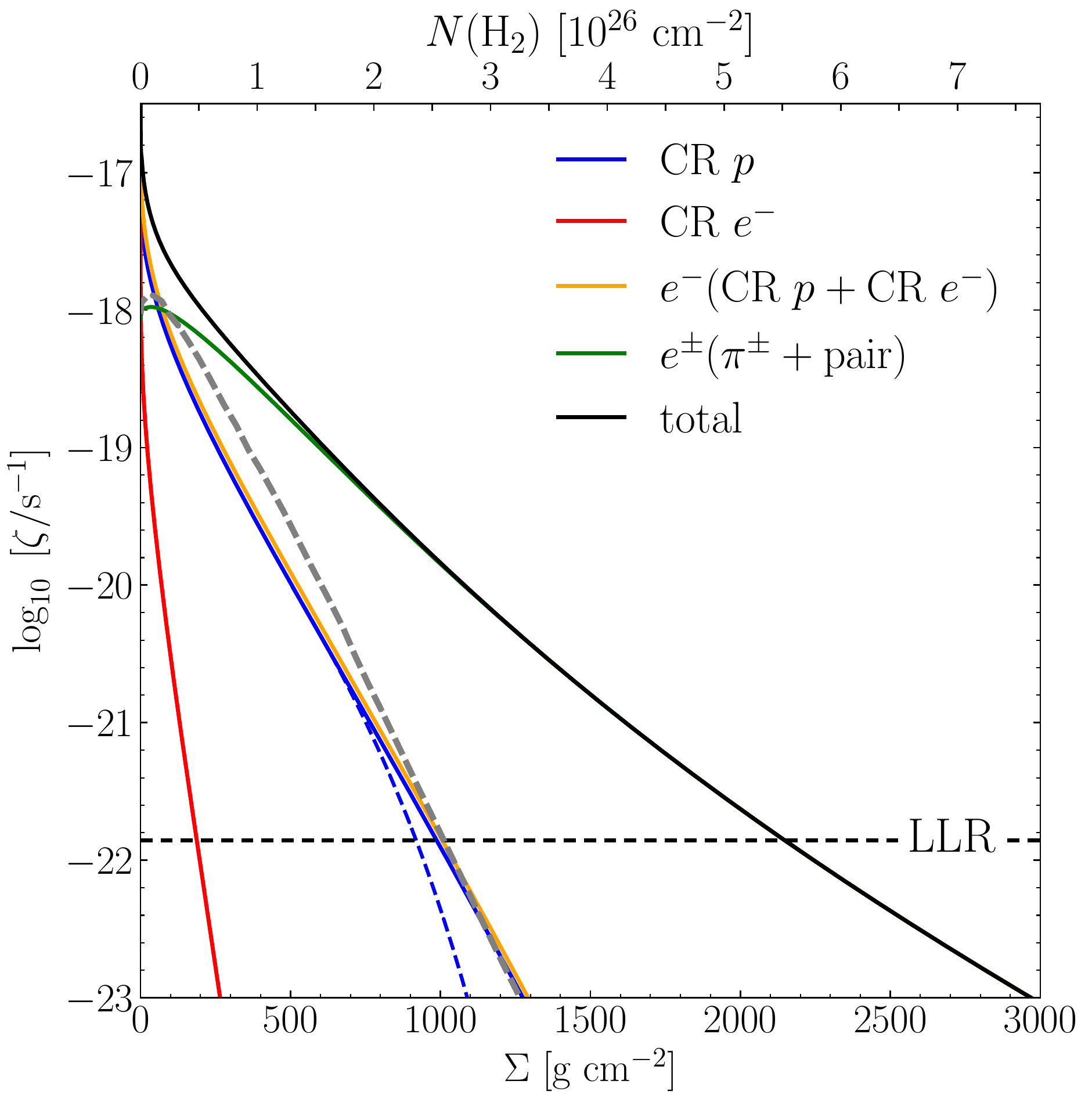}}
\caption{Ionisation rate
per $\hh$ due to primary and secondary CR species, $\zeta$, plotted vs. the surface density $\Sigma$ (bottom scale) and the
column density $N$ (top scale). The {\em black line} shows the total ionisation rate. Partial contributions to $\zeta$ include
ionisation due to primary CR protons and electrons ({\em blue and red lines}, respectively), ionisation due to  
secondary electrons created by primary CRs ({\em orange line}), and ionisation due to 
electrons and positrons created by
charged pion decay and pair production ({\em green line}). The {\em blue dashed line} shows the proton contribution
calculated with the CSDA approach. The horizontal {\em dashed line} at $1.4\times10^{-22}$~s$^{-1}$ indicates the total
ionisation rate set by long-lived radioactive nuclei (LLR). For comparison, we also plot the total ionisation rate per
$\hh$ obtained by Umebayashi \& Nakano~(\cite{un81}) ({\em grey dashed line}). The total rate of the electron production,
$\zeta_e$, is approximately $1.11\zeta$.} \label{crion}
\end{center}
\end{figure}

Figure~\ref{crion} shows the total ionisation rate and partial contributions from various species.  For $\Sigma$ below the transition surface density,
$\Sigma_{\rm tr}\approx130$~g~cm$^{-2}$, the ionisation is mainly due to CR protons (and their secondary electrons),
while at higher surface densities the contribution of electron-positron pairs produced by photon decay   becomes progressively dominant. At $\Sigma\gtrsim600$~g~cm$^{-2}$, pairs fully determine the ionisation---their contribution is about a factor of 10 larger than that of CR protons---that is,
the ionisation is no longer
affected by the magnetic field and hence is controlled by the
line of sight, rather than the effective column density. 
We note that in previous studies the CR ionisation rate was computed as a function of the line-of-sight surface density (e.g. Umebayashi \& Nakano~\cite{un81}). 
As long as the ionisation rate is dominated by charged particles ($\Sigma\lesssim\Sigma_{\rm tr}$), 
the relevant quantity is the effective 
surface density seen by CRs
moving along magnetic field lines ($\Sigma_{\rm eff}$).
Depending on the magnetic field configuration
(see e.g. Padovani et al.~\cite{phg13}),
the latter is generally larger or much larger than the 
line-of-sight surface density ($\Sigma_{\rm los}$).
The ionisation rate can be approximated by the following expression:
\be
\zeta\simeq\left\{ \begin{array}{ll}\label{zefflos}
 \zeta(\Sigma_{\rm eff}) & \textrm{for $\Sigma_{\rm eff}\le\Sigma_{\rm tr}$}\\
 \langle\zeta(\Sigma_{\rm tr}+\Sigma_{\rm los})\rangle & \textrm{for $\Sigma_{\rm eff}>\Sigma_{\rm tr}$}
  \end{array} \right.\,,
\ee
where brackets
denote averaging over all the directions
from the transition surface towards a given position.
An application of this formula is given in Sect.~\ref{Ionisation_CD}.

We note that direct photoionisation is always 
negligible since the photoionisation cross section rapidly decreases with increasing energy (see Appendix~\ref{phionsigma}).
Furthermore, because of a (partially) diffusive transport of CR protons (see Sect.~\ref{propagationp}), their
contribution, $\zeta_p(\Sigma)$, is described by a Gaussian curve at large $\Sigma$. In Fig.~\ref{crion} we compare this
(solid blue) curve with $\zeta_p(\Sigma)$ calculated with the pure CSDA approach.\footnote{The CSDA curve in
Fig.~\ref{crion} decreases more steeply than the diffusive curve at larger $\Sigma$, since the CSDA formalism implies the
existence of a certain terminal column density beyond which CR protons cannot penetrate. The latter directly follows from
Eq.~(\ref{range_0}) taking into account that $L_p(E)$ at high energies increases faster than linearly
(Eq.~\ref{pionlosses}).} The results essentially coincide up to $\Sigma\approx600$~g~cm$^{-2}$, where the
contribution of electrons and positrons dominates.

Our important conclusion is that at large surface densities the ionisation is determined by 
electron-positron pairs 
and that the ionisation rate is not exponential anymore. The reason why the electron and positron ionisation dominates at
large $\Sigma$ can be inferred from Fig.~\ref{photons} by comparing the behaviour of the photon,
electron, and positron
spectra. We see that the photon spectrum is entirely due to BS generated by $e^\pm$, whereas the
electron and positron spectra are entirely due to the pairs produced by photons. This indicates that the feedback loop
$e^\pm\xrightarrow{\rm BS}\gamma\xrightarrow{\rm pair}e^\pm$ in Fig.~\ref{diagram} starts playing a crucial role in the
ionisation process. Physically, this is because photons are able to propagate far from the source. Therefore at large
$\Sigma$, where primary CRs are completely attenuated and ionisation is due to secondary particles, photons provide the
only mechanism of efficient transport and ionisation (by generating pairs).

Our results are substantially different from those obtained by Umebayashi \& Nakano~(\cite{un81}). They found the total
ionisation rate decreases exponentially with a characteristic attenuation scale of about 115~g~cm$^{-2}$ for
$100~\mathrm{g~cm}^{-2}\lesssim\Sigma\lesssim500~\mathrm{g~cm}^{-2}$, and about 96~g~cm$^{-2}$ at larger surface densities.
Conversely, we find a characteristic scale that continuously increases with surface density, from $\approx112$~g~cm$^{-2}$
to $\approx285$~g~cm$^{-2}$ in the range $100~\mathrm{g~cm}^{-2}\lesssim\Sigma\lesssim2100~\mathrm{g~cm}^{-2}$, within an
error lower than 10\%. This difference is because Umebayashi \& Nakano~(\cite{un81}) treated proton-proton
collisions above the threshold for pion production as catastrophic losses, and described high-energy Compton scattering with
the CSDA approach. In addition, we consider the presence of heavy elements both in the IS CR flux and in the
target medium, and perform the pitch-angle-averaging for CR protons in the CSDA regime.\footnote{ At lower column
densities, $\zeta_{p}(N)$ can be derived analytically. In Appendix~\ref{ion_low} we present a typical solution for the local
proton spectrum and show that the resulting $\zeta_{p}(N)$, Eq.~(\ref{zeta_low}), is described by a power-law dependence in
the column density range ${\rm10^{19}~cm^{-2}\le}~N~{\rm\le10^{25}~cm^{-2}}$.}

For computational purposes (e.g. numerical simulations and chemical models), in Appendix~\ref{polyfit} we give a polynomial
fit of $\zeta(N)$ valid in the range ${\rm10^{19}~cm^{-2}\le}~N~{\rm\le10^{27}~cm^{-2}}$. We point out that the total
rate of electron production, $\zeta_{e}$, is slightly larger than $\zeta$ because it also includes ionisation
reactions of CRs with He (see Table~1 in PGG09; contributions of heavier species of the medium are
negligible). For $f_{\rm He}$ from Table~\ref{tabwilms}, we find
\be \zeta_{e}(N)\approx1.11\zeta(N)\,, \ee
with an accuracy of 1\% for the same range of $N$.

\section{Applications}
\label{discussion}

The ionisation fraction is a fundamental quantity for the dynamics of the IS gas, in particular during
the earliest stages of star formation, from the collapse of a molecular cloud core to the formation of an
accretion disc. Before the formation of a protostar, CR ionisation regulates the
degree of coupling between gas and magnetic field in the densest parts of a cloud core,
setting the timescale of magnetic field diffusion
(see e.g. Pinto et al.~\cite{pgb08}) and controlling the amount of magnetic braking of collapsing rotating envelopes
(Galli et al.~\cite{gl06}; Li et al.~\cite{lk16}).

In our previous studies on CR propagation (PGG09; Padovani \& Galli~\cite{pg11,pg13}; Padovani et al.~\cite{pgg13,phg13,pg14}), we neglected
the contribution of electron-positron pairs generated by
photon decay and that of relativistic protons and electrons. While this approximation
is appropriate for diffuse or dense molecular clouds, it becomes invalid
at the high values of column densities characteristic of circumstellar discs, where
CR ionisation is dominated by both relativistic and secondary particles.
As shown in Sect.~\ref{ionisation}, at surface densities larger than $\Sigma\approx 130$~g~cm$^{-2}$
the ionisation rate due to CR protons quickly becomes 
negligible, but pair production maintains $\zeta$ larger than the LLR ionisation threshold up
to $\Sigma\approx2100$~g~cm$^{-2}$.

The ionisation rate in a circumstellar disc varies considerably with radius and vertical height above the disc mid-plane,
and is produced by several ionising agents, such as Galactic CRs, accelerated particles and X-rays from the central star,
and short- or long-lived radioactive elements mixed in the gas,  whose relative importance depends on the specific
conditions. A careful determination of the ionisation fraction in such environment is crucial to assess the efficiency of the
magnetorotational instability and the existence of the so-called dead zones with respect to mass and angular momentum transport (Gammie~\cite{g96}).

In the following subsections we concentrate on the effects of Galactic CRs penetrating in discs around protostars and young
stars, and limit our analysis to the disc mid-plane, where terrestrial planets are likely to form. Our objective is to
quantify the dependence of the CR ionisation rate on the disc physical characteristics (surface density and magnetic field
profiles) rather than providing an exhaustive analysis of all possible sources of ionisation. To this goal, we consider
several idealised models of magnetised discs around young stars (from Shu et al.~\cite{sg07}) characterised by a power-law
behaviour of the relevant properties, including the benchmark case of the (unmagnetised) minimum mass solar nebula (MMSN;
Hayashi~\cite{hay81}). We also examine the effect of CR modulation by a stellar wind (Cleeves et al.~\cite{ca13a}) and
calculate the ionisation rate due to stellar particles (Rab et al.~\cite{rg17}) in the particular case of a disc around a
T-Tauri star at a radius of 1~AU (postponing a more detailed analysis to a future study).

We stress that our results (and results by Umebayashi \& Nakano~\cite{un81}) do not apply to the regions of the disc
dominated by MHD turbulence. In our analysis we only include the effects of large-scale magnetic fields threading the disc,
ignoring the scattering and diffusion of CRs due to the turbulent magnetic field. For a recent calculation of CR propagation
in this regime, see Rodgers-Lee et al.~(\cite{rl17}).

\subsection{Ionisation in magnetised circumstellar discs by Galactic CRs}
\label{Ionisation_CD}

We derive the CR ionisation rate in the mid-plane of circumstellar discs with various surface density distributions
and magnetic field profiles. We choose the models defined by Shu et al.~(\cite{sg07}), representative of low- and high-mass
protostars (LMP and HMP), T Tauri stars (TT), and FU Orionis stars (FU Ori). These disc models are characterised by the mass
of the central star, the accretion rate, and the disc age (see Table~2 in Shu et al.~\cite{sg07}). They are described by
power-law surface density profiles and mid-plane magnetic field strength scaling with radius $\varpi$ as
\be \Sigma_{\rm disc}=\Sigma_0\left(\frac{\varpi}{100~{\rm AU}}\right)^{-3/4}, \qquad
B_z=B_0\left(\frac{\varpi}{100~{\rm AU}}\right)^{-11/8}, \ee
with $\Sigma_0=1.36$, 8.42, 33.6, and 59.5~g~cm$^{-2}$, and $B_0=9.07$, 8.70, 55.2, and 164~mG for TT, LMP, FU Ori, and HMP,
respectively\footnote{In this Section, $\Sigma_{\rm disc}$ denotes the vertically integrated total surface density of a
disc.}. 
In contrast, the unmagnetised MMSN model has a surface density
of 1.7 g~cm$^{-2}$ at $\varpi=100$~AU and a surface density profile proportional to $\varpi^{-3/2}$  (Hayashi~\cite{hay81}).

\begin{figure}[!h]
\begin{center}
\resizebox{\hsize}{!}{\includegraphics{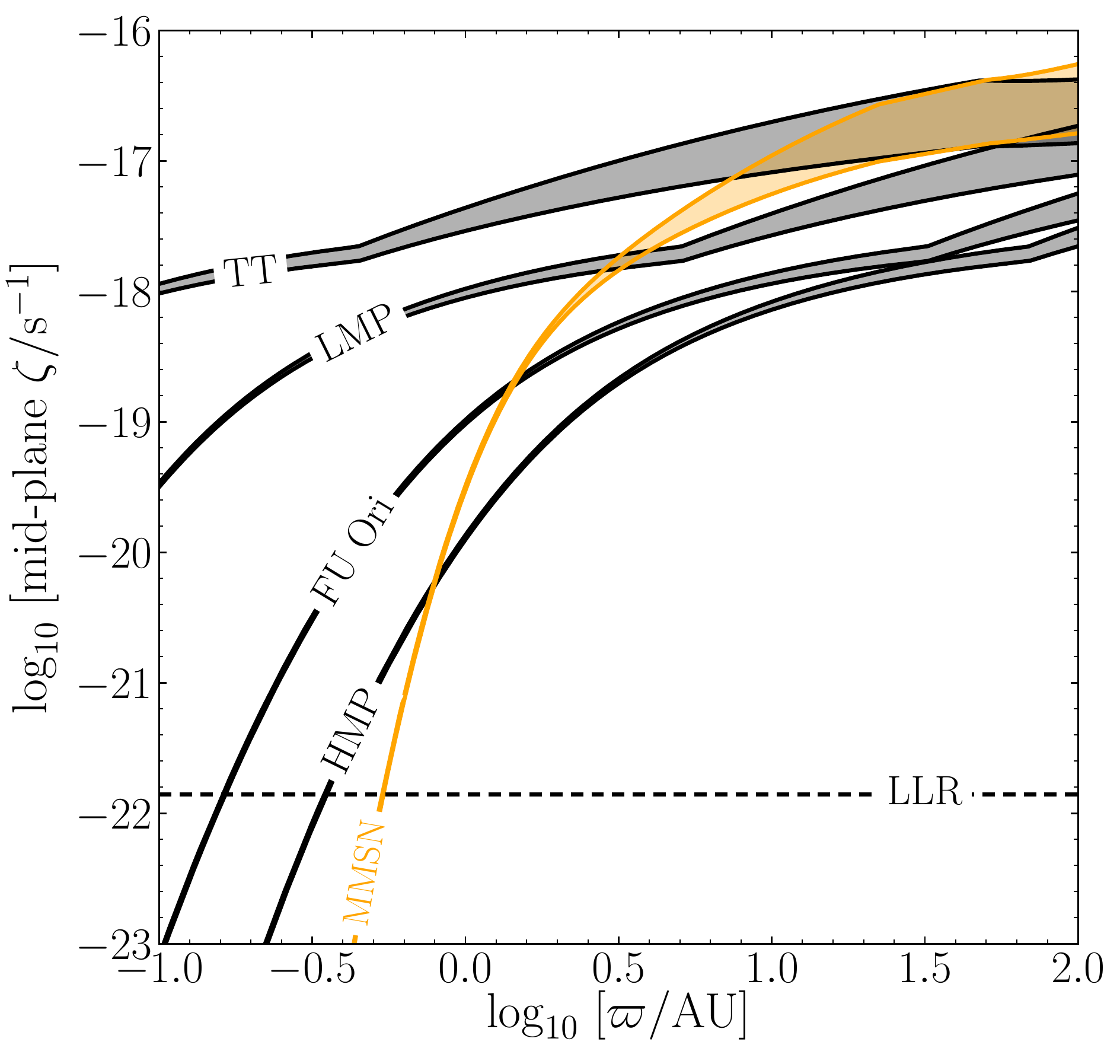}}
\caption{Mid-plane CR ionisation rate per $\hh$, $\zeta$,
plotted against radius $\varpi$ in the TT, LMP,
FU Ori, and HMP models from Shu et al.~(\cite{sg07})
together with the standard (unmagnetised)  MMSN model from Hayashi~(\cite{hay81}).
The upper and lower borders of the {\em shaded areas}
correspond to the unmodulated Galactic CR spectra ${\mathscr H}$ and ${\mathscr L}$, respectively.
The kinks seen at the level of $\zeta\approx2\times10^{-18}$~s$^{-1}$
occur at the transition surface density $\Sigma_{\rm tr}=130$~g~cm$^{-2}$ in Eq.~(\ref{zefflos}).
The horizontal {\em dotted line} at $1.4\times10^{-22}$~s$^{-1}$ shows the value of the ionisation rate set by LLR. } \label{accdiscs}
\end{center}
\end{figure}

The effective surface density crossed by a CR propagating along a magnetic field line, inclined with respect to the disc
plane, is $\Sigma_{\rm eff}=\Sigma_{\rm disc}/\cos\psi$, where $\cos\psi=B_{z}/B$. In the models by Shu et al.~(\cite{sg07}), the
factor $1/\cos\psi$ is independent of the disc type and approximately equal to $3.3$. 
As we noted in Sect.~\ref{ionisation}, below $\Sigma_{\rm tr}\approx130$~g~cm$^{-2}$ the ionisation
is controlled by the effective surface density measured along magnetic field lines, while
above $\approx600$~g~cm$^{-2}$ 
becomes
independent of the magnetic field configuration and hence
is determined by the line-of-sight surface density.
Equation~(\ref{zefflos}) can be used to compute the ionisation rate in the disc mid-plane as a function 
of radius $\varpi$.
Figure~\ref{accdiscs} shows $\zeta$ 
for the various models.
The mid-plane CR ionisation rate becomes dominated by LLRs inside $\varpi\approx0.5$~AU, $0.3$~AU, and
$0.1$~AU for MMSN, HMP, and FU Ori, respectively. Only for the most evolved TT discs and for LMP,
$\zeta$ is always larger than $\zeta_{\rm LLR}$. We note that the typical age of a TT
disc is much larger than the
half-life of SLRs (e.g. $^{26}$Al has a half-life of 0.74~Myr; Umebayashi \& Nakano~\cite{un09}), whose contribution to the
ionisation is therefore negligible.

The results shown in Fig.~\ref{accdiscs} illustrate the contribution of unshielded
Galactic CRs to the ionisation in the
disc mid-plane (see Sect.~\ref{T-Tauri} for the effects of stellar modulation). Among other sources of ionisation in discs
around young stars, X-rays play an important role (Igea \& Glassgold~\cite{ig99}), but the value of the X-ray ionisation
rate $\zeta_X$ at high column densities is difficult to compute because of the limitations in the Monte Carlo
scattering calculations. In practice, $\zeta_X$ becomes uncertain above $\Sigma_{\rm disc} \approx
70$~g~cm$^{-2}$ (Ercolano \& Glassgold~\cite{eg13}).
For the MMSN model, this range corresponds to radii smaller than  $\approx 8$~AU, where Galactic CRs -- if not strongly
affected by the stellar wind -- would mostly dominate the ionisation in the mid-plane.

\subsection{Ionisation in discs around T-Tauri stars}
\label{T-Tauri}

\begin{figure}[!h]
\begin{center}
\resizebox{\hsize}{!}{\includegraphics{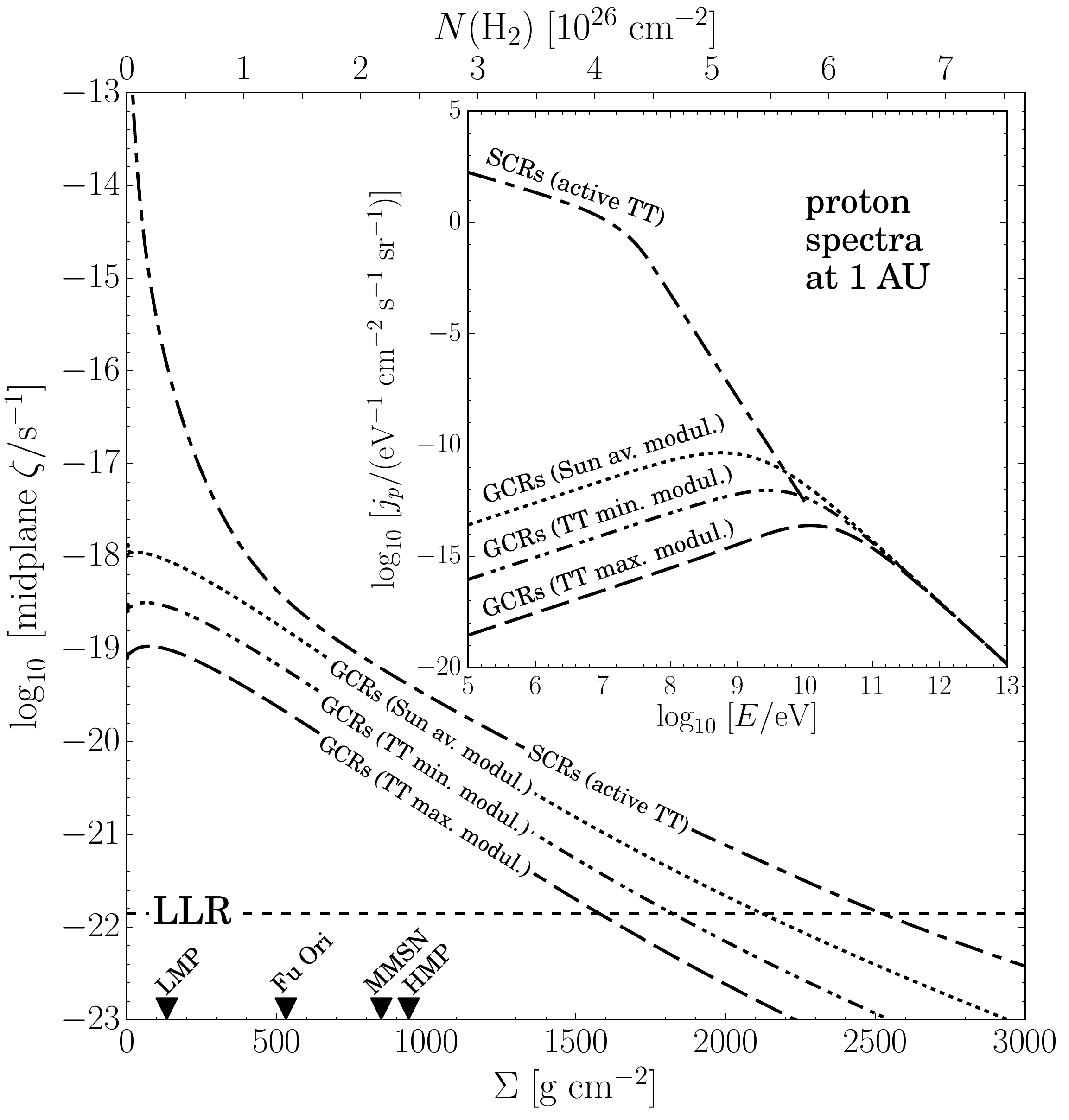}}
\caption{Mid-plane CR ionisation rate per $\hh$, $\zeta$, vs. the surface density $\Sigma$
(bottom scale) and column density $N$ (top scale) plotted for several proton spectra (at a distance of
1~AU from the central star), as shown in the inset. Galactic CRs with solar mean modulation ({\em dotted line}); minimum and
maximum modulation by a TT wind ({\em dash-dot-dotted} and {\em long dashed line}, respectively); and stellar CRs
from an active TT star ({\em dash-dotted line}). The horizontal {\em dashed line} shows the ionisation rate set by
LLR. For reference, {\em black triangles} on the horizontal axis indicate the values of the
mid-plane disc density ($\Sigma_{\rm disc}/2$) at 1~AU for the LMP, FU Ori, MMSN, and HMP models
(134, 532, 850, and 942 g~cm$^{-2}$, respectively).
}
\label{fig12}
\end{center}
\end{figure}

Low-energy Galactic CRs may be prevented from penetrating an extended heliosphere (or ``T Tauriosphere'') surrounding a
young star (Cleeves et al.~\cite{ca13a}). Unfortunately, little can be said about the extent and shape of this region of CR
exclusion other than scaling up the properties of the Sun's heliosphere. Cleeves et al.~(\cite{ca13a}) suggested that the
T Tauriosphere may well surround the entire disc. However, the energies of CR particles mostly responsible for the
ionisation at column densities above $\approx 100$~g~cm$^{-2}$ exceed a few GeV. The effect of the modulation by the stellar
wind at these energies is uncertain. For TT stars, Cleeves et al.~(\cite{ca13a}) estimated values for the
modulation potential $\phi$ at a distance of 1~AU in the range $\phi=4.8-18$~GeV, leading to a
reduction of the CR flux at $E=10$~GeV by a factor of $\approx 6$ and $100$, respectively ($\phi$ is an unknown
function of distance, which could be determined from detailed magnetospheric models). 
Moreover, the presence of a
young active star may lead to increased ionisation rate, at least in the regions closest to the star because of thermal
ionisation (Nakano \& Umebayashi et al.~\cite{nu88}), particle emission in flares and/or coronal mass ejection shock waves
(Reames~\cite{r13,r15}), or via locally accelerated CRs in circumstellar and jet shocks (Padovani et
al.~\cite{phm15,pmh16,pmh17}).

We apply our model of CR propagation to estimate the ionisation produced at a distance of 1~AU from the
protostar by two different input proton spectra: a spectrum of Galactic protons modulated by TT stellar
winds (Usoskin et al.~\cite{ua05}; Cleeves et al.~\cite{ca13a}), and an enhanced flux of stellar protons
generated by flares in an active TT star (Feigelson et al.~\cite{fg02}; Rab et al.~\cite{rg17})\footnote{We note,
however, that the rectilinear propagation of stellar protons {assumed by Rab et al.~(\cite{rg17})} may lead to an
overestimate of the proton flux incident on the disc at 1~AU (Fraschetti et al.~\cite{fd18}).}. We find significant
differences with respect to previous results:

({\em i}\/) Fig.~\ref{fig12} shows the CR ionisation rate for maximum and minimum modulation by a TT stellar wind
at 1~AU, corresponding to $\phi=18$~GeV and $\phi=4.8$~GeV, respectively, and labelled by ``GCRs (max. modul.)'' and ``GCRs
(min. modul.)''. For completeness, the figure also shows the ionisation rate for the Galactic CR flux modulated by
an average solar wind ($\phi=1$~GeV) labelled ``GCRs (Sun. av. modul.)''. To facilitate the comparison with
previous studies, we do not take into account inclination of the magnetic field lines with respect to the disc plane
(considered in Sect.~\ref{Ionisation_CD}). Compared to Cleeves et al.~(\cite{ca13a}), our result for the minimum
modulation model is larger by a factor of $\approx30$ at $\Sigma\lesssim100$~g~cm$^{-2}$, while above
$\Sigma\approx1200$~g~cm$^{-2}$ it decreases much more abruptly. The difference is even more dramatic for the maximum
modulation model. We find an ionisation rate that is larger by a factor of $\approx260$ at $\Sigma\lesssim100$~g~cm$^{-2}$
and is decreasing faster above $\Sigma\approx1400$~g~cm$^{-2}$. The discrepancy at low surface densities is largely
because of our inclusion of the process of electron-positron pair creation by photon decay and also because of the adoption of
the relativistic behaviour of the ionisation cross section for protons. The faster decrease of our results at high
surface densities is caused by losses due to heavy elements in the medium (see Sect.~\ref{energylosses}). It is
noteworthy that, in contrast to cases of unmodulated Galactic CRs, the ionisation rate for the minimum and maximum
modulation is almost entirely due to relativistic protons, propagating diffusively (see Sect.~\ref{propagationp}). Hence,
for the modulated cases we do not perform the pitch-angle averaging.

({\em ii}\/) For the proton flux generated in a TT flare, labelled in Fig.~\ref{fig12} by ``SCRs (active TT)'',
our results for up to $\Sigma\approx300$~g~cm$^{-2}$ agree with those obtained by Rab et al.~(\cite{rg17}) within
5\%.
At higher surface densities the ionisation rate computed with our model decreases slowly, since 
electron-positron pairs
increase the ionisation by orders of magnitude. It is important to remark that is still unclear what fraction of CRs
generated in a flare event can be channeled into the disc through magnetic field lines, without crossing the turbulent zone,
and what part may follow open field lines perpendicular to the disc (Shu et al.~\cite{sn00}; Feigelson et al.~\cite{fg02}).

\section{Conclusions}
\label{conclusions}

The main result of this paper is the characterisation of the CR ionisation rate at high column densities. In
particular, we showed how the CSDA fails to describe the CR proton propagation above the energy threshold for pion
production ($E^{\pi}=280$~MeV). In fact, when a CR proton interacts with a local proton to create a pion, the energy
loss of the CR proton is not small anymore and there is also a certain degree of scattering. These two effects go against
the main assumptions of the CSDA, namely the infinitesimal energy loss and the conservation of  pitch angles. Furthermore,
we carefully described the production of secondary particles, focussing on the propagation of photons created by neutral pion
decay and by secondary electrons and positrons through BS. In previous studies, photon Compton losses have been
treated by assuming CSDA, but we demonstrated that for this process it is crucial to use a diffusion equation. An accurate
description of photon propagation is essential, since the electron and positron fluxes depend on the photon flux. 

It is important to stress the main difference between the secondary particle showers that we consider here and the CR
air showers in the Earth atmosphere, where the decay length of muons or pions is comparable to the
scale height of the atmosphere. A big effort has been made to explain Earth air showers observed with, for example the Auger
Observatory detectors and with Imaging Atmospheric Cherenkov Telescopes such as H.E.S.S. and MAGIC. Uncertainties on
the hadronic cascades are the main source of error in the determination of the composition of ultra-high-energy CRs. These
errors can be reduced only through detailed air shower simulations and comparisons with Large Hadron Collider and CR data
(Pierog~\cite{pie17}). For circumstellar discs the situation is the opposite:\ one can assume that muons and pions immediately decay and, hence, safely consider photons and secondary 
electrons and positrons
as direct products of the CR interaction with the local medium.

Our main conclusions are
\begin{itemize}
\item Interstellar CR protons and particles produced by the secondary mechanisms penetrate much farther
    inside a circumstellar disc with respect to what has been calculated in previous studies.
    As a consequence, the CR ionisation rate remains above the value set by LLRs up to
    the surface density of $\approx2100$~g~cm$^{-2}$;
\item Primary CRs are completely attenuated at high surface densities, and (secondary) photons are then the
    only species that propagate and efficiently create electron-positron pairs. In turn, these pairs  produce
    efficient ionisation and, through BS, create the next generation of photons, which leads to the
    ionisation feedback loop \mbox{$\gamma\xrightarrow{\rm pair}e^\pm \xrightarrow{\rm BS}\gamma$};
\item The total ionisation rate $\zeta$ as a function of $\Sigma$ (or $N$) cannot be described by an exponential
    attenuation law. In fact, the attenuation scale {continuously increases with surface density from
    $\approx112$~g~cm$^{-2}$ to $\approx285$~g~cm$^{-2}$ in the range $\mathrm{100~g~cm^{-2}}\lesssim
    \Sigma\lesssim\mathrm{2100~g~cm^{-2}}$}. Our results are considerably different from the exponential attenuation by
    Umebayashi \& Nakano~(\cite{un81}). The difference is because of a number of improvements in our model: $(i)$ collisions
    of CR protons with energies above the pion production threshold and photon Compton scattering are treated as
    a diffusion process; $(ii)$ we account for the presence of heavy elements both in the CR flux and in the target
    medium; and $(iii)$ we perform the averaging over initial pitch angles of CR protons in the CSDA regime.
\item The ionisation rate for $\Sigma\lesssim130$~g~cm$^{-2}$ is determined by CR protons (and their secondary
    electrons), while for $\Sigma\gtrsim600$~g~cm$^{-2}$
    it is completely controlled by secondary photons that  create
    electron-positron pairs (producing local ionisation). Therefore, $\zeta(\Sigma)$ is a function of the 
    effective surface density (measured along the magnetic field lines) at $\Sigma\lesssim  130$~g~cm$^{-2}$
    and of the line-of-sight surface density at $\Sigma\gtrsim600$~g~cm$^{-2}$, since the photon propagation is unaffected by the
    magnetic field.
\item We show that $\zeta(N)$ can be described analytically by a power-law dependence in the range
    ${\rm10^{19}~cm^{-2}\le}~N~{\rm\le10^{25}~cm^{-2}}$; for practical purposes, we also give a polynomial fit in
    the whole range ${\rm10^{19}~cm^{-2}\le}~N~{\rm\le10^{27}~cm^{-2}}$ (where $N$ is related to $\Sigma$ by
    Eq.~\ref{NSigma}). The implementation of this fitting formula in numerical simulations and astrochemical codes is
    straightforward.

\end{itemize}

We applied our method to the propagation of CRs in magnetised circumstellar discs around young stars (Shu et
al.~\cite{sg07}) where the ionisation fraction (which depends on the CR ionisation rate) is a key parameter that controls
the coupling of the gas to disc magnetic field, the efficiency of the MRI instability, and the occurrence of dead zones. Our
results can be easily incorporated in disc models together with the effects of other sources of ionisation (most
importantly, X-rays) not considered in our analysis. However, a better understanding of the process of exclusion of Galactic
CRs by stellar winds is needed for disc surface densities below $\approx 150$~g~cm$^{-2}$, where the CR ionisation is
largely due to protons with energies below $\approx 1$~GeV (which are strongly affected by stellar modulation).

Finally we checked how 
the secondary CR particles, composition of the medium, and averaging over the initial pitch angles affect the
ionisation rate. Two different input proton spectra were considered: an IS CR proton flux modulated by TT
stellar winds (Cleeves et al.~\cite{ca13a}) and a local stellar proton flux generated in a flare event of an active TT
(Feigelson et al.~\cite{fg02}; Rab et al.~\cite{rg17}). We found as follows:\begin{itemize}
\item While stellar winds are able to devoid the IS spectrum of low-energy protons (below $\approx 1$~GeV),
    the high-energy part of the spectrum is responsible for the production of 
    electron-positron pairs through photon
    decay. The pair ionisation {(along with the adoption of the relativistic ionisation cross section for
    protons)} keeps the ionisation rate at $\Sigma\lesssim100$~g~cm$^{-2}$ much larger than previously calculated by a
    factor of 30 and 260 for the minimum and maximum modulation model, respectively.
    Furthermore, our value of $\zeta(\Sigma)$ calculated for these models decreases much faster at
    $\Sigma\gtrsim1200$~g~cm$^{-2}$ and $\gtrsim1400$~g~cm$^{-2}$, respectively, because of the larger energy losses
    (determined by the medium composition).
\item For typical ages of TT discs, the ionisation by SLRs -- if initially present -- is negligible. The
    ionisation plateau is set by LLRs, and the CR ionisation dominates up to
        $\approx1580$~g~cm$^{-2}$ and $\approx1820$~g~cm$^{-2}$ for the maximum and minimum modulation model,
        respectively.
\item For ionisation due to stellar particles created in a TT flare, our results are comparable to previous
    calculations below $\approx300$~g~cm$^{-2}$ within 5\%, while at higher surface densities 
    electron-positron pairs increase the ionisation rate by orders of magnitude.
\end{itemize}

In this paper we developed a model of ionisation at high densities, above a few g~cm$^{-2}$, particularly relevant
for the inner regions of collapsing clouds and circumstellar discs. We calculated dependencies $\zeta(\Sigma)$, representing
several characteristic energy spectra of CRs. Apart from an extreme (and also poorly constrained) case of ionisation due to
enhanced flux of stellar protons, the obtained dependencies can be considered as fairly universal and applicable to any
relevant environment. The principal limitation of our results is that they cannot be generally used to describe ionisation
in regions dominated by MHD turbulence, which may lead to a diffusive transport of CRs (essentially dependent on properties
of the turbulence). We plan to systematically investigate the effect of MHD turbulence in a separate paper.

\begin{acknowledgements}
 The authors wish to thank the referee, Christopher McKee, for his careful reading of
the manuscript and
insightful comments that considerably helped to improve the
paper and Elena Amato for valuable discussions. MP acknowledges funding from the European Unions Horizon 2020 research and
innovation programme under the Marie Sk\l{}odowska-Curie grant agreement No 664931. PC acknowledges support from the
European Research Council (ERC; project PALs 320620).
\end{acknowledgements}

\appendix

\section{Energy loss functions}

In this Appendix we discuss the individual contributions to the energy loss functions for (primary and secondary) CR
particles interacting with a local IS medium composed by a mixture of hydrogen, helium, and heavier species
according to Wilms et al.~(\cite{wa00}), see Table~\ref{tabwilms}.

\begin{table}[]
\caption{Assumed composition of IS medium. Number of electrons ($Z$), mass number ($A_{Z}$), and abundance with
respect to the total number of particles ($f_{Z}$). 
}
\begin{center}
\begin{tabular}{cccc}
\hline\hline
species & $Z$ & $A_{Z}$ & $f_{Z}^{(a)}$ \\
\hline
$\hh$ & 2 & 2 & $8.35\times10^{-1}$ \\
He    & 2 & 4 & $1.63\times10^{-1}$ \\
C     & 6 & 12& $4.01\times10^{-4}$ \\
N     & 7 & 14& $1.27\times10^{-4}$ \\
O     & 8 & 16& $8.19\times10^{-4}$ \\
Ne    & 10& 20& $1.46\times10^{-4}$ \\
Na    & 11& 23& $2.41\times10^{-6}$ \\
Mg    & 12& 24& $4.19\times10^{-5}$ \\
Al    & 13& 27& $3.57\times10^{-6}$ \\
Si    & 14& 28& $3.11\times10^{-5}$ \\
P     & 15& 31& $4.39\times10^{-7}$ \\
S     & 16& 32& $2.05\times10^{-5}$ \\
Cl    & 17& 35& $2.21\times10^{-7}$ \\
Ar    & 18& 40& $4.29\times10^{-6}$ \\
Ca    & 20& 40& $2.64\times10^{-6}$ \\
Ti    & 22& 48& $1.08\times10^{-7}$ \\
Cr    & 24& 52& $5.41\times10^{-7}$ \\
Mn    & 25& 55& $3.66\times10^{-7}$ \\
Fe    & 26& 56& $4.49\times10^{-5}$ \\
Co    & 27& 59& $1.39\times10^{-7}$ \\
Ni    & 28& 59& $1.87\times10^{-6}$ \\
\hline
\end{tabular}
\end{center}
\footnotesize{$(a)$ computed assuming the IS medium composition by Wilms et al.~(\cite{wa00}).}
\label{tabwilms}
\end{table}%

\subsection{Protons}\label{Lproton}

The main contribution to $L_{p}$ at low energies is due to ionisation losses that are proportional to the atomic number
$Z$ of the target species {(Bethe-Bloch formula, see Hayakawa~\cite{h69})}, so that $L_{p,Z}^{\rm ion}=ZL_{p,\h}^{\rm ion}$, while for molecular hydrogen
$L_{p,\hh}^{\rm ion}=2L_{p,\h}^{\rm ion}$. The total ionisation loss function reads
\be\label{Lpion}
L_{p}^{\rm ion}(E)=\left(2f_{\hh}+\sum_{Z\geq2}f_{Z}Z\right)L_{p,\h}^{\rm ion}(E)=%
\varepsilon^{\rm ion}L_{p,\h}^{\rm ion}(E)\,,
\ee
where $\varepsilon^{\rm ion}=2.01$. 

At higher energies, above a threshold
$E^{\pi}=280$~MeV, we add energy losses due to pion production, as given by Schlickeiser~(\cite{s02}) and Krakau \&
Schlickeiser~(\cite{ks15a}),
\be\label{pionlosses}
L^{\pi}_{p,Z}(E)\approx 2.57\times10^{-17}\frac{A_{Z}^{0.79}}{\beta}\left(\frac{E}{\rm GeV}\right)^{1.28}
\left(\frac{E+E^{\rm as}}{\rm GeV}\right)^{-0.2}~{\rm eV~cm^{2}}
\ee
where $\beta=v/c$ is the ratio between the proton speed and the speed of light, and the asymptotic
energy $E^{\rm as}=200$~GeV. The factor $A_{Z}^{0.79}$ is a phenomenological correction to the pion production cross
section for heavier target species (Geist~\cite{g91}).
Pion losses become dominant for $E\gtrsim 1$~GeV, fully determining the propagation of high-energy CRs at high column
densities.
The total pion production loss function reads
\be\label{Lpi} L_{p}^{\pi}(E)=\left(2f_{\hh}+\sum_{Z\geq2}f_{Z}A_{Z}^{0.79}\right)L_{p,\h}^{\pi}(E)=
\varepsilon^{\pi}L_{p,\h}^{\pi}(E)\,, \ee
where $\varepsilon^{\pi}=2.17$.

\subsection{Electrons and positrons}\label{Lelectron}

Ionisation losses for electrons have the same correction factor as protons, $L_{e}^{\rm ion}(E)=\varepsilon^{\rm
ion}L_{e,\h}^{\rm ion}(E)$, see Eq.~(\ref{Lpion}).

BS losses dominate at $E\gtrsim100$~MeV. We take into account that $L_{e,\hh}^{\rm BS}=2L_{e,\h}^{\rm BS}$
and that the differential BS cross section is proportional to $Z(Z+1)$;  see Appendix~\ref{bremsssigma}. This
yields
\be\label{LBS}
L_{e}^{\rm BS}(E)=\left(2f_{\hh}+\sum_{Z\geq2}f_{Z}\frac{Z(Z+1)}{2}\right)L_{e,\h}^{\rm BS}(E)=%
\varepsilon^{\rm BS}L_{e,\h}^{\rm BS}(E)\,,
\ee
where $\varepsilon^{\rm BS}=2.24$.

Synchrotron losses dominate at energies above $E^{\rm syn}\approx 1$~TeV and do not depend on the composition. Following
Schlickeiser~(\cite{s02}), $L^{\rm syn}_e(E)$ is
\be\label{Lsyn} L^{\rm syn}_e(E)\approx 5.0\times10^{-14}\left(\frac{E}{\rm TeV}\right)^{2}\, \mathrm{eV~cm^{2}}\,, \ee
where we have assumed a relation between the magnetic field strength, $B$, and the gas number
density, $n$, given by Crutcher~(\cite{c12})
\be\label{Bn} B=B_{0}\left(\frac{n}{n_{0}}\right)^{\kappa}\,, \ee
with $B_{0}\approx 10~\mu$G, $n_{0}=150$~cm$^{-3}$ and $\kappa\approx 0.5$--0.7. The value of $\kappa$ recommended by
Crutcher~(\cite{c12}) is $\kappa=0.65$, but we assume $\kappa=0.5$ (Nakano et al. \cite{nn02}; Zhao et al. \cite{z16}) to
benefit from the removal of the dependence of $L^{\rm syn}_{e}$ on $n$. 

For positrons we adopt the same total loss function as for electrons, and therefore use the same notation $L_e$ for
both species.

\subsection{Photons}\label{Lphoton}

Photoionisation and pair production are catastrophic processes. Their loss functions are proportional to the
corresponding cross sections $\sigma^{\rm PI}$ (see Appendix~\ref{phionsigma}) and $\sigma^{\rm pair}$ (see
Appendix~\ref{bremsssigma}). Compton effect is a continuous loss process; its energy loss function (Eq.~\ref{loss_cont}) is
determined by the Compton differential cross section, $\ud\sigma^{\rm C}/\ud E_{e}$ (Eq.~\ref{dsCdEe}), and the maximum
kinetic energy transferred to the recoiling electron, $E_{e}^{\rm max}$ (Eq.~\ref{Ee_max}).

Since Compton losses are proportional to $Z$, the correction due to heavy elements is the same as
ionisation losses. Hence,
$\ud\sigma^{\rm C}/\ud E_{e}=\varepsilon^{\rm C}\ud\sigma^{\rm C}_{\h}/\ud E_{e}$ with $\varepsilon^{\rm C}=2.01$ (see
Appendix~\ref{Csigma}).

\section{Cross sections}
\label{x-sections}

\subsection{Elastic proton-nucleus collisions}\label{mtsigmaeqs}

The differential cross section for proton-proton collision in the centre-of-mass reference system is given in Jackson \&
Blatt~(\cite{jb50}) as the sum of three terms: one for elastic (Coulomb) scattering, $\ud\sigma^{\rm E}/\ud\Omega$, one for
nuclear scattering, $\ud\sigma^{\rm N}/\ud\Omega$, and an interference term that can be neglected. In a normalised form, the
first two terms are written%
\begin{eqnarray}
\frac{\ud\tilde\sigma^{\rm E}(E,\vartheta)}{\ud\Omega}&=&\csc^4\left(\frac{\vartheta}2\right)+\sec^4\left(\frac{\vartheta}2\right)
\\\nonumber
&&      -\csc^2\left(\frac{\vartheta}2\right)\sec^2\left(\frac{\vartheta}2\right)%
        \cos\left[2\frac{\beta}{\alpha}\ln\tan\left(\frac{\vartheta}2\right)\right],\\
\frac{\ud\tilde\sigma^{\rm N}(E,\vartheta)}{\ud\Omega}&=&4\frac{\beta^2}{\alpha^2}\sin^2\delta_{0}\,,
\end{eqnarray}
where $\alpha=e^2/\hbar c$ is the fine-structure constant, $\beta=v/c$ is determined by the relative velocity of the two
protons, $\delta_{0}(E,\vartheta)$ is the nuclear phase shift (Breit et al.~\cite{bt39}; Jackson \& Blatt~\cite{jb50}), and
$\vartheta$ is the scattering angle. The differential cross section in the centre-of-mass reference frame is then
\begin{equation}
\frac{\ud\sigma^{pp}(E,\vartheta)}{\ud\Omega}=\frac{r_p^2}{\beta^4}\left(\frac{\ud\tilde\sigma^{\rm E}}{\ud\vartheta}+
        \frac{\ud\tilde\sigma^{\rm N}}{\ud\vartheta}\right)\,,
\end{equation}
where $r_{p}=e^2/m_pc^2$ is the classical proton radius.
The momentum transfer cross section is written
\begin{equation}
\sigma_{\rm MT}^{pp}(E)=2\pi\int_0^\pi\frac{\ud\sigma^{pp}}{\ud\Omega}(1-\cos\vartheta)\sin\vartheta\,\ud\vartheta\,.
\end{equation}
To account for the collisions between CR protons and target heavy nuclei, $\sigma_{\rm MT}^{pp}$ has to be multiplied
by the correction factor $\xi$ given by
\be\label{xifac} \xi=f_{\hh}+\sum_{Z\geq2}Z^{2}\frac{A_{Z}}{A_{Z}+1}f_{Z}=1.49\,. \ee
The factor $A_{Z}/(A_{Z}+1)$ on the right-hand side accounts for the efficiency of momentum transfer from a CR proton to a nucleus with
the mass number $A_Z$ (Landau \& Lifshitz~\cite{LL69}); in the first term we take into account that for collisions with
H$_2$ the cross section is $2\sigma_{\rm MT}^{pp}$.

\subsection{Photoionisation}\label{phionsigma}

The photoionisation cross section, $\sigma^{\rm PI}$, accounting for medium composition, is
written as%
\be \sigma^{\rm PI}(E)=f_{\hh}\sigma^{\rm PI}_{\hh}(E)+\sum_{Z\geq2}f_{Z}\sigma^{\rm PI}_{Z}(E)\,. \ee
The cross sections for different species on the right-hand side are given by Yeh \& Lindau~(\cite{yl85}) and
Yeh~(\cite{y93})\footnote{See also \url{https://vuo.elettra.eu/services/elements/WebElements.html}.}, and are matched to the
asymptotic behaviour (Draine~\cite{d11}). Being expressed in terms of $\alpha$ and the Bohr radius $a_0=\hbar^2/m_ee^2$, the
asymptotic cross section is written
\be \sigma^{\rm PI}(E)=\frac{2^{8}}{3Z^{2}}\alpha\pi a_{0}^{2}\left(\frac{E}{Z^{2}I_{\h}}\right)^{-3.5} \ee
where $I_{\h}=13.6$~eV is the ionisation energy of atomic hydrogen (valid for energies much larger than $Z^{2}I_{\h}$). 

\subsection{Bremsstrahlung and pair production}
\label{bremsssigma}

The differential cross section for BS of electrons on hydrogen atom is given by the Bethe-Heitler
formula
\begin{equation}\label{BHbremss}
\frac{\ud \sigma^{\rm BS}_{\h}(E_e,E_\gamma)}{\ud E_\gamma}=\frac{\alpha r_e^2}{E_\gamma} \left\{[1+\left(1-x
\right)^{2}]\phi_1(\Delta)-\frac{2}{3}\left(1-x\right)\phi_2(\Delta)\right\},
\end{equation}
where $x=E_\gamma/(E_e+m_{e}c^{2})$ and
\begin{equation}
\Delta=\left(\frac{m_e c^2}{4\alpha E_\gamma}\right)\frac{x^2}{(1-x)}.
\end{equation}
The functions $\phi_1(\Delta)$ and $\phi_2(\Delta)$ are tabulated in Table~2 of Blumenthal \& Gould~(1970). A simple
analytical fit is
written as\begin{equation}
\phi_{1,2}(\Delta)\approx 8\left[\ln\left(\frac{1}{2\alpha(1+\Delta)}\right)+\frac{c_{1,2}-\Delta}{1+2\Delta}\right],
\end{equation}
where $c_1=3/2$ and $c_2=4/3$. This formula has the correct behaviour both for $\Delta\ll 1$ and $\Delta\gg 1$.

For heavier species, the differential cross section for BS is a factor $Z(Z+1)$ larger than that of atomic hydrogen.
This factor comes from the fact that BS takes place in the nuclear Coulomb field and in
the field of atomic electrons. Consequently, BS losses are proportional to $Z(Z+1)$ rather than $Z^{2}$
(Wheeler \& Lamb~\cite{wl39}; Hayakawa~\cite{h69}). The differential BS cross section for H$_{2}$ is a factor of
2 larger than that of atomic hydrogen (Gould~\cite{g69}).
Equation~(\ref{BHbremss}) holds for relativistic energies. For lower energies we
used the parameterisations given by Koch \& Motz~(\cite{km59}) and Sacher \& Sch\"onfelder~(\cite{ss84}). We note that the
differential cross section is divergent for $E_\gamma\rightarrow 0$.

The differential cross section for pair production by a photon in the field of a nucleus is closely related to that for
BS, since the Feynman diagrams are variants of one another. For H nuclei,
\begin{equation}\label{d_sigma_pair}
\frac{\ud \sigma^{\rm pair}_{\h}(E_{\gamma},E_e)}{\ud E_e}=\frac{\alpha r_e^2}{E_\gamma} \left\{[y^2+(1-y)^2]\phi_1(\delta)
+\frac{2}{3}y(1-y)\phi_2(\delta)\right\},
\end{equation}
where $y=(E_e+m_{e}c^{2})/E_\gamma$ and \be \delta=\left(\frac{m_ec^2}{\alpha E_\gamma}\right)\frac{1}{y(1-y)}. \ee The differential
cross section is clearly symmetric for $y \leftrightarrow 1-y$.
The total pair production cross section,
\begin{eqnarray}
\sigma^{\rm pair}_{\h}(E_\gamma) &=& \int_0^{E_\gamma} \frac{\ud \sigma^{\rm pair}_{\h}}{\ud E_e}\,\ud E_e  \label{sigma_pair}\\
&=& \alpha r_e^2\int_0^1\left\{[y^2+(1-y)^2]\phi_1(\delta)
+\frac{2}{3}y(1-y)\phi_2(\delta)\right\}\,\ud y,\nonumber
\end{eqnarray}
has the asymptotic limit for $E_\gamma\rightarrow\infty$,
\begin{equation}
\sigma^{\rm pair}_{\h}(E_\gamma)\rightarrow \alpha r_e^2\left[\frac{2}{3}\phi_1(0)+\frac{1}{9}\phi_2(0)\right]=20.6~\mbox{mb}\,.
\end{equation}
As for BS, it holds $\sigma_{\hh}^{\rm pair}=2\sigma_{\h}^{\rm pair}$ and $\sigma_{Z}^{\rm
pair}=Z(Z+1)\sigma_{\h}^{\rm pair}$.

\subsection{Compton scattering}\label{Csigma}

The differential cross section of Compton scattering for atomic hydrogen, expressed in terms of the incident photon
energy $E_\gamma$ and scattering angle $\vartheta$, is given by the Klein-Nishina formula  
(Hayakawa~\cite{h69})
\begin{equation}\label{KN}
\frac{\ud\sigma^{\rm C}_{\h}(E_\gamma,\vartheta)}{\ud\Omega}=\frac{1}{2}r_{e}^{2}\left(\frac{x'}{x}\right)^{2}
\left(\frac{x'}{x}+\frac{x}{x'}-\sin^{2}\vartheta\right)\,,
\end{equation}
where $r_{e}=e^2/m_ec^2$ is the classical electron radius. Here $x=E_{\gamma}/(m_{e}c^{2})$ and
$x'=E'_{\gamma}/(m_{e}c^{2})$ are normalised energies before and after scattering, related by
\begin{equation}\label{Compton_theta}
\frac{1}{x'}-\frac{1}{x}=1-\cos\vartheta.
\end{equation}
The kinetic energy transferred to the recoiling electron is $E_e=E_\gamma-E'_\gamma$; its maximum value,
\begin{equation}\label{Ee_max}
E_e^{\rm max}=\frac{2x^2}{1+2x}m_ec^2,
\end{equation}
corresponds to $\vartheta=\pi$. The differential cross section is straightforwardly derived from
Eq.~(\ref{KN}): substituting $\vartheta(E_\gamma,E_e)$ and using $\ud\cos\vartheta/\ud E_e=-(E_\gamma-E_e)^{-2}$, which follows
from Eq.~(\ref{Compton_theta}), we get
\begin{equation}\label{dsCdEe}
\frac{\ud\sigma^{\rm C}_{\h}(E_\gamma,E_e)}{\ud E_{e}}=\frac{2\pi}{(E_\gamma-E_e)^2}
\frac{\ud\sigma^{\rm C}_{\h}(E_\gamma,E_e)}{\ud\Omega}\,.
\end{equation}
The cross section for Compton scattering for atomic hydrogen is obtained by integrating Eq.~(\ref{KN}) over the
solid angle,
\begin{eqnarray}\label{Comptonsigma}
\sigma^{\rm C}_{\h}(E_\gamma)&=&\frac34\sigma_{\rm T}\left\{\frac{1+x}{x^{2}}\left[\frac{2(1+x)}{1+2x}
-\frac{\ln(1+2x)}{x}\right]\right.\\\nonumber
&&+\left.\frac{\ln(1+2x)}{2x}-\frac{1+3x}{(1+2x)^{2}}\right\}\,,
\end{eqnarray}
where $\sigma_{\rm T}=\frac83\pi r_e^2$ is the Thomson cross section. Asymptotically, $\sigma^{\rm C}_{\h}\approx\sigma_{\rm
T}$ for $x\ll1$ and $\sigma^{\rm C}_{\h}\approx\frac38\sigma_{\rm T}\ln(2x)/x$ for $x\gg1$. We have $\sigma_{\hh}^{\rm
C}=2\sigma_{\h}^{\rm C}$ and $\sigma^{\rm C}_{Z}\propto Z$, so the total Compton cross section is given by
\be\label{sigCtot} \sigma^{\rm C}=\left(2f_{\hh}+\sum_{Z\geq2}f_{Z}Z\right)\sigma_{\h}^{\rm C}%
=2.01\sigma_{\h}^{\rm C}\,. \ee

The momentum transfer cross section for atomic hydrogen,
\begin{equation}\label{sigmamtC}
\sigma_{\rm \h,MT}^{\rm C}(E)=2\pi\int_{0}^{\pi}\frac{\ud\sigma^{\rm
C}_{\h}}{\ud\Omega}(1-\cos\vartheta)\sin\vartheta\,\ud\vartheta
\end{equation}
has the following analytic form for $x>10^{-3}$:
\begin{eqnarray}
\sigma_{\rm \h,MT}^{\rm C}(E_\gamma)&=&\frac34\sigma_{\rm T}\left[\frac{2}{(1+2x)^{2}}+\frac{2x-\ln(1+2x)}{x^{2}}\right.\\\nonumber
&&-\left.\frac{2x(3+x)-(3+4x)\ln(1+2x)}{x^{4}}\right].
\end{eqnarray}
For $x<10^{-3}$ it tends to $\sigma_{\rm \h,MT}^{\rm C}=2\sigma_{\rm T}$. Similar to Eq.~(\ref{sigCtot}), we obtain
$\sigma_{\rm MT}^{\rm C}\approx2\sigma_{\h,\rm MT}^{\rm C}$\,. 

\section{Matching CSDA and the diffusive regimes for protons}\label{match}

The two regimes of propagation must be matched at the transition energy $E^{\rm tr}$ (see lower panel of Fig.~\ref{sigmt}):
for $E\geq E^{\rm tr}$ the diffusion solution is given by Eq.~(\ref{diffsol}), which yields the matching spectrum
$j_p(E^{\rm tr},N)$ for the CSDA regime operating at lower energies. For brevity, below we omit particle indices and
introduce the auxiliary function $R(E,E_0)\equiv R(E_0)-R(E)$, determined by the range function, Eq.~(\ref{range_0}). Two
solutions are possible in the CSDA regime, depending on how $N$ compares to the transition column density $N_{\rm
tr}=R(0,E^{\rm tr})$:

$(i)~N\le N^{\rm tr}$: the energy range is divided into two parts, $E\le E_*$ and $E_*<E< E^{\rm tr}$, with $E_*$ determined
from $R(E_*,E^{\rm tr})=N$. For $E\le E_*$ the local flux is given by the attenuated IS spectrum,
\begin{equation}\label{E<}
j(E,N)=j_{\rm IS}(E_{0})\frac{L(E_{0})}{L(E)}\,,
\end{equation}
with $E_{0}$ from $R(E,E_0)=N$, whereas for $E_*<E< E^{\rm tr}$ it is governed by the matching spectrum,
\begin{equation}\label{E>}
j(E,N) = j(E^{\rm tr},N-\Delta N)\frac{L(E^{\rm tr})}{L(E)}\,,
\end{equation}
with $\Delta N=R(E,E^{\rm tr})$. We note that for $E=E_*$ we have $E_0=E^{\rm tr}$ in Eq.~(\ref{E<}) and $\Delta N=N$ in
Eq.~(\ref{E>}). Since $j(E,0)=j_{\rm IS}(E)$, the solution is continuous at $E=E_*$.

$(ii)~N>N^{\rm tr}$: the IS spectrum is completely attenuated, so Eq.~(\ref{E>}) is valid for all $E< E^{\rm tr}$.

\section{Losses due to elastic proton-nucleus collisions}
\label{elastic}

Elastic collisions of CR protons with nuclei of the medium are accompanied by energy exchange. As this process is most
efficient for particles of equal mass, let us examine the effect of proton-proton collisions and consider for simplicity
the CSDA regime governed by Eq.~(\ref{transport_eq}). The energy exchange leads to a sink term $-\sigma^{pp}(E)j_p(E,N)$
(to be added to the right-hand side), describing a depopulation of CR energy state $E$ due to elastic collisions with hydrogen nuclei.
There is also a  source term that consists of two contributions: $S_p^{(1)}$, due to depopulation of higher energy CR
state $E+E'$ (accompanied by exchange of energy $E'$ with a hydrogen nucleus), and $S_p^{(2)}$, due to hydrogen nuclei that
acquire energy $E$ after collisions with CR protons;
it is naturally assumed that such collisions result in dissociation of
molecular hydrogen. By introducing the differential cross section of proton-proton collisions, $\ud \sigma^{pp}/\ud \Delta
E$, which is a function of CR energy $E$ and energy exchange $\Delta E$, we have
\begin{equation}
S_p^{(1)}(E,N)=\int_0^\infty\frac{\ud \sigma^{pp}(E+E',E')}{\ud \Delta E}j_p(E+E',N)\;\ud E',
\end{equation}
while $S_p^{(2)}(E,N)$ is given by the same expression with arguments $(E+E',E)$ for the cross section.

The inclusion of these additional terms in Eq.~(\ref{transport_eq}) generally results in a complicated integro-differential
equation for CR protons. These terms (negligible for non-relativistic energies, where ionisation losses dominate) may play a
role for relativistic protons. The interaction with the medium is then mostly due to nuclear scattering, which is characterised
by hard-sphere-like cross section (see upper panel of Fig.~\ref{sigmt}). In this case $\ud\sigma^{pp}/\ud \Delta E\approx
\sigma^{pp}/E$, i.e. the differential cross section does not depend on $\Delta E$ and is determined by a constant
$\sigma^{pp}$ (equal to $\sigma^{pp}_{\rm MT}\approx 3$~mb). By substituting the resulting source and sink terms in
Eq.~(\ref{transport_eq}), we obtain the following transport equation for CR protons ($\mu=1$):
\begin{equation}
\frac{\partial j_p}{\partial N}-\frac{\partial}{dE}\left(L_pj_p\right)
=2\sigma^{pp}\int_0^\infty\frac{j_p(E+E')}{E+E'}\;\ud E'-\sigma^{pp}j_p.
\end{equation}
An approximate solution of this equation for high energies can be factorised,
\begin{equation}\label{j1}
j_p(E,N)\approx e^{-\sigma^{pp}_{\rm eff}N}j_p'(E,N),
\end{equation}
where $j_p'(E,N)$ is a solution of (homogeneous) Eq.~(\ref{transport_eq}) and $\sigma^{pp}_{\rm eff}$ is an unknown
effective cross section, describing the cumulative effect of elastic nuclear collisions and depending on the form of
$j_{p}^{\rm IS}(E)$. To obtain $\sigma^{pp}_{\rm eff}$, we notice that the loss function in the high-energy regime is
dominated by the pion production and, according to Eq.~(\ref{pionlosses}), can be roughly approximated by
$L_p^{\pi}(E)\approx \delta_\pi\sigma^\pi E$, where $\sigma^\pi\approx 32$~mb is the pion production cross section
(neglecting a weak logarithmic energy dependence) and $\delta_\pi\approx 0.3$ is the energy fraction lost in a single
collision. Assuming $j_p^{\rm IS}\propto E^{-\nu}$, we get
\begin{equation}\label{j2}
j'(E,N)=e^{-(\nu-1)\delta_\pi\sigma^\pi N}j_{p}^{\rm IS}(E)
\end{equation}
and $\sigma^{pp}_{\rm eff}=(1-2/\nu)\sigma^{pp}$. Since $\delta_\pi\sigma^\pi$ is of the order of $\sigma^{pp}$ and
$\nu\approx 2.7$ (relativistic part of $j_{p}^{\rm IS}$), we conclude that the argument of the exponential in Eq.~(\ref{j1})
is much smaller than that in Eq.~(\ref{j2}), i.e. the contribution of elastic collisions of CR protons can be safely
neglected.

\section{Ionisation by CR protons at low column densities}
\label{ion_low}

Consider CR ionisation at relatively low $N$, where ionisation is still the main loss mechanism. Our numerical results show
that, for model $\mathscr{L}$ and $\mathscr{H}$ of IS proton spectra $j_p^{\rm IS}(E)$, the contribution of CR
electrons to $\zeta(N)$ can be neglected at $N\gtrsim 10^{19}$~cm$^{-2}$ and $N\gtrsim 3\times10^{21}$~cm$^{-2}$,
respectively. Therefore, starting from these column densities we are only interested in the propagation of CR protons.

The upper panel of Fig.~\ref{Lfunc} shows that ionisation dominates losses for non-relativistic protons, and for
$10^5$~eV~$\lesssim E\lesssim 5\times10^8$~eV the loss function is very well approximated by a single power-law dependence,
\begin{equation}\label{L_ion}
L_p(E)=AE^{-s},
\end{equation}
where $A=1.77\times10^{-10}$~eV~cm$^{2}$ and $s\approx 0.82$ (energy is in eV). The propagation and attenuation of such
protons occurs in the CSDA regime and is governed by Eq.~(\ref{transport_eq}). Furthermore, from Fig.~\ref{fig1bis} we infer
that these energies correspond to a very broad range of column densities, 
$10^{19}$~cm$^{-2}\lesssim N\lesssim
10^{25}$~cm$^{-2}$. By substituting Eq.~(\ref{L_ion}) in Eq.~(\ref{transport_eq}), we derive the following general solution
valid for these $N$:
\begin{equation}
j_p(E,N)=E^s\Psi\left[E^{1+s}+(1+s)AN\right],
\end{equation}
where the function $\Psi(x)$ is determined by matching $j_p(E,0)$ with the IS spectrum $j_p^{\rm IS}(E)$. For
instance, for a power-law spectrum $j_p^{\rm IS}\propto E^{-\nu}$ we get
\begin{equation}\label{spectrum_solution}
j_p(E,N)=j_p^{\rm IS}(E)\left[1+(1+s)AN/E^{1+s}\right]^{-\frac{\nu+s}{1+s}}.
\end{equation}
With the derived local spectrum, it is straightforward to obtain $\zeta(N)$. We substitute Eq.~(\ref{spectrum_solution}) in
Eq.~(\ref{crionint}) and notice that the cross section of ionisation by non-relativistic protons obeys a power-law scaling
for $E\gtrsim10^5$~eV, $\sigma_p^{\rm ion}\propto E^{-b}$ with $b\approx s$, i.e. $L_p(E)/\sigma_p^{\rm ion}(E)$ is
practically independent of $E$ and hence $\Phi_p(E)\approx $~const in Eq.~(\ref{crionint}). Then integration over $E$
yields the following dependence:
\begin{equation}\label{zeta_low}
\zeta(N)=c_1+c_2N^{-q},
\end{equation}
where $q=(\nu+b-1)/(1+s)$ and $c_{1,2}$ are constants. Equation~(\ref{zeta_low}) is obtained assuming $1+s-b>0$, and is
valid as long as $q>0$, i.e. for $\nu>1-b$. For $b\approx s\approx 0.82$ we obtain $q\approx 0.55\nu-0.10$, which is valid for
$\nu\gtrsim0.2$. We note that here $\nu$ represents the non-relativistic part of $j_p^{\rm IS}(E)$, e.g. for model ${\mathscr
H}$ ($\nu=0.8$) we have $q\approx 0.34$. The lower bound of applicability of Eq.~(\ref{zeta_low}) is determined by the
actual form of $j_p^{\rm IS}(E)$, as mentioned above, while the upper bound is $N\approx 10^{25}$~cm$^{-2}$ (or
$\Sigma\approx 40$~g~cm$^{-2}$).

\section{Polynomial fit of the CR ionisation rate at any column density}\label{polyfit}

\begin{figure}[!htb]
\begin{center}
\resizebox{\hsize}{!}{\includegraphics{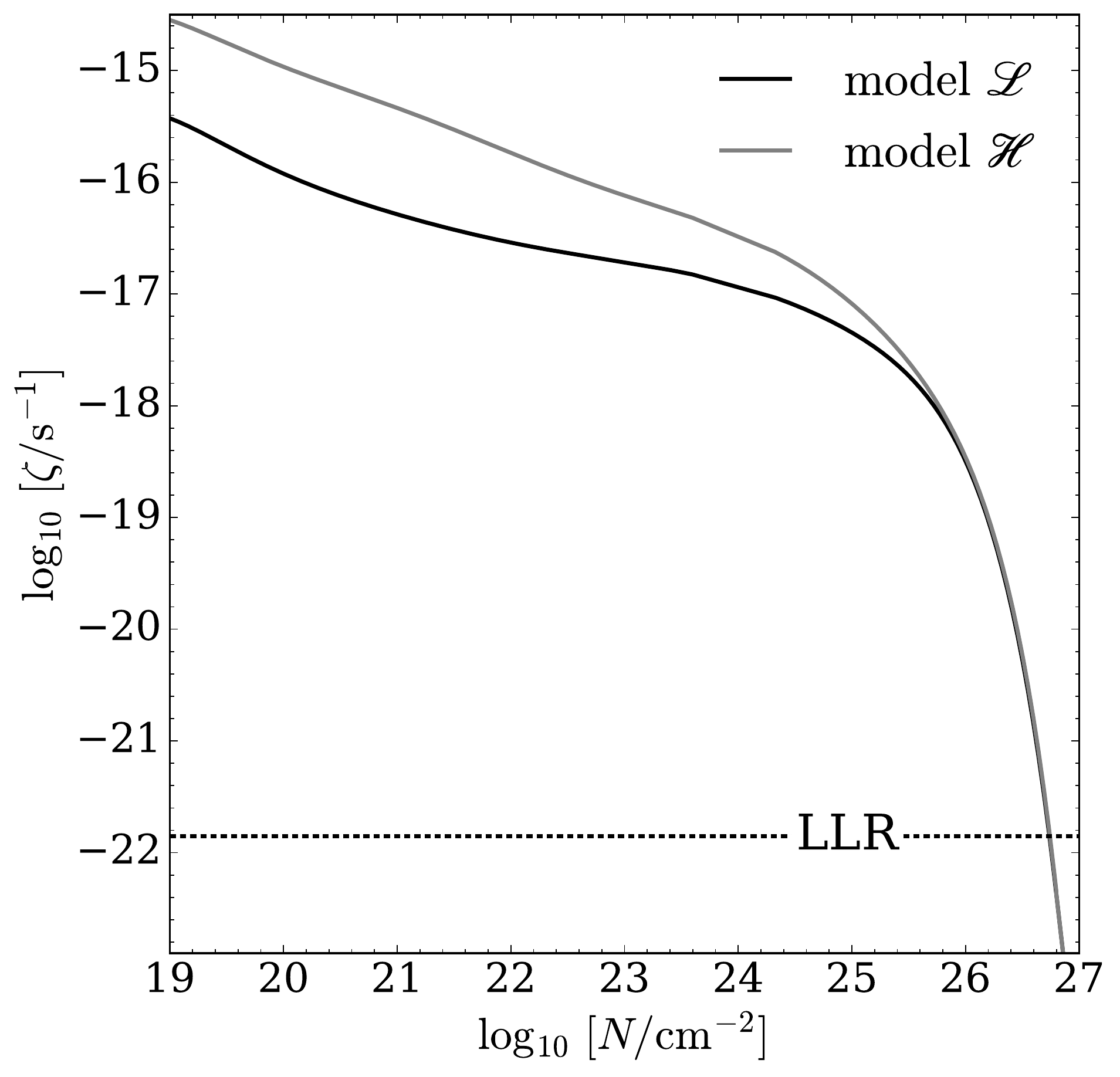}}
\caption{
Total CR ionisation rate $\zeta$ per $\hh$ due to primary and secondary CR species plotted vs. the column density
$N$. The horizontal {\em dashed line} at $1.4\times10^{-22}$~s$^{-1}$ indicates the total ionisation rate set by LLR. Models $\mathscr{L}$ (black) and $\mathscr{H}$ (grey) are 
described by Eq.~(\ref{jis}).}
\label{zetavsN}
\end{center}
\end{figure}

For practical purposes, the total CR ionisation rate (of $\hh$) can be parameterised with the following fitting
formula:
\be\label{eqpolyfit} \log_{10}\frac{\zeta}{\rm s^{-1}}=\sum_{k\geq0}c_{k}\log_{10}^{k}\frac{N}{\rm cm^{-2}}\,. \ee
It is applicable for column densities ${\rm10^{19}~cm^{-2}\le}~N~{\rm\le10^{27}~cm^{-2}}$ with a maximum error of 6\% and
an average accuracy of 2\%. Table~\ref{coeffpolyfit} gives two sets of coefficients, $c_{k}$, for both models
$\mathscr{L}$ and $\mathscr{H}$, since at low column densities the ionisation rate depends on the low-energy CR proton
spectrum. Figure~\ref{zetavsN} shows a log-log plot of $\zeta$ versus $N$ for the two models.

According to Eq.~(\ref{zefflos}), $\zeta(N)$ is a function of the effective column density as long
as $\Sigma\lesssim 130$~g~cm$^{-2}$ ($N\lesssim 3\times10^{25}$~cm$^{-2}$). The excess over this transition value should be
calculated as the line-of-sight column density.
\begin{table}[htp]
\caption{Coefficients $c_{k}$ of the polynomial fit, Eq.~(\ref{eqpolyfit}),
for two models of IS proton spectra (see Sect.~\ref{spectra}).}
\begin{center}
\begin{tabular}{ccc}
\hline\hline
$k$ & model $\mathscr{L}$ & model $\mathscr{H}$\\
\hline
0 &           $-3.331056497233\times10^6$    & $\phantom{-}1.001098610761\times10^7$\\
1 & $\phantom{-}1.207744586503\times10^6$    &           $-4.231294690194\times10^6$\\
2 &           $-1.913914106234\times10^5$    & $\phantom{-}7.921914432011\times10^5$\\
3 & $\phantom{-}1.731822350618\times10^4$    &           $-8.623677095423\times10^4$\\
4 &           $-9.790557206178\times10^2$    & $\phantom{-}6.015889127529\times10^3$\\
5 & $\phantom{-}3.543830893824\times10^1$    &           $-2.789238383353\times10^2$\\
6 &           $-8.034869454520\times10^{-1}$ & $\phantom{-}8.595814402406\times10^0$\\
7 & $\phantom{-}1.048808593086\times10^{-2}$ &           $-1.698029737474\times10^{-1}$\\
8 &           $-6.188760100997\times10^{-5}$ & $\phantom{-}1.951179287567\times10^{-3}$\\
9 & $\phantom{-}3.122820990797\times10^{-8}$ &           $-9.937499546711\times10^{-6}$\\
\hline
\end{tabular}
\end{center}
\label{coeffpolyfit}
\end{table}%

\end{document}